\documentclass[11pt,a4paper
]{article}
\usepackage{amsthm,amsfonts,amsmath,amscd,amssymb}
\usepackage{latexsym}
\usepackage{euscript}
\usepackage{enumitem}
\usepackage{graphicx}
\usepackage{tikz}
\usepackage{caption}
\usepackage{subcaption}
\usepackage{cite}
\usepackage[utf8]{inputenc}
\usepackage[english]{babel}
\usepackage{xcolor}

\usepackage{hyperref}
\usepackage{etoolbox}
\usepackage{forloop}

\hypersetup{colorlinks=true}
\numberwithin{equation}{section}
\numberwithin{figure}{section}
\numberwithin{table}{section}

\newcommand{\key}[1]{\par\vskip 0.5em\small{\slshape Key words:\/} #1}

\newcommand{\msc}[1]{\par\vskip 0.5em\small{\slshape Mathematical Subject Classification 2020:\/} #1}

\newcommand\arXiv[1]{\url{https://arxiv.org/abs/#1}}

\theoremstyle{plain}
\newtheorem{theorem}{Theorem}

\newtheorem{lemma}[theorem]{Lemma}
\theoremstyle{definition}
\newtheorem{definition}[theorem]{Definition}
\theoremstyle{remark}

\usepackage{bm}

\title
{ASYMPTOTICS OF CORRELATORS OF SPARSE BIPARTITE RANDOM GRAPHS}

\author{Valentin Vengerovsky\footnote{B. Verkin Institute for Low Temperature Physics and Engineering of the National Academy of Sciences of Ukraine, 47 Nauky Ave., Kharkiv, 61103, Ukraine. E-mail: \href{mailto:vengerovsky@ilt.kharkov.ua}{vengerovsky@ilt.kharkov.ua}}}

\begin{document}
\maketitle



\newcommand{\Compl}{\mathbb{C}}
\newcommand{\R}{\mathbb{R}}
\newcommand{\Z}{\mathbb{Z}}

\newcommand{\conseq}{\Rightarrow}

\newcommand{\Mat}{\mathrm{Mat}}
\newcommand{\conj}[1]{\overline{#1}}
\newcommand{\transp}[1]{#1^T}
\newcommand{\rev}[1]{\dual{#1}}
\newcommand{\dual}[1]{#1^R}
\newcommand{\cConjScl}[1]{\bar{#1}}
\newcommand{\cConjMat}[1]{#1^*}
\newcommand{\aConjScl}[1]{#1^*}
\newcommand{\aConjMat}[1]{#1^+}
\newcommand{\ds}[1]{\check{#1}}

\newcommand{\setcharf}[1]{\mathbbm{1}_{#1}}
\newcommand{\charfp}{\psi}
\newcommand{\asKer}[2]{K\left(#1, #2\right)}
\newcommand{\CF}{\mathsf{f}}

\newcommand{\Gin}{\mathrm{Gin}}
\newcommand{\ens}{M_n}
\newcommand{\sclA}{y}
\newcommand{\matA}{Y}

\newcommand{\cSclGin}{x}
\newcommand{\cMatGin}{X}
\newcommand{\cMatPos}{\mathcal{Z}}
\newcommand{\cSclN}{t}
\newcommand{\cVecN}{\bm{\cSclN}}
\newcommand{\cSclA}{q}
\newcommand{\cMatA}{Q}
\newcommand{\cSetMatA}{\bm{Q}}
\newcommand{\cVecGI}{\bm{h}}
\newcommand{\cSclB}{a}

\newcommand{\aSclA}{\xi}
\newcommand{\aVecA}{\bm{\aSclA}}
\newcommand{\aMatA}{\Xi}
\newcommand{\aSetMatA}{\mathbf{\aMatA}}
\newcommand{\aSclB}{\phi}
\newcommand{\aVecB}{\bm{\aSclB}}
\newcommand{\aMatB}{\Phi}
\newcommand{\aSclBt}{\varphi}
\newcommand{\aVecBt}{\aVecB}
\newcommand{\aSclC}{\theta}
\newcommand{\aVecC}{\bm{\aSclC}}
\newcommand{\aMatC}{\Theta}
\newcommand{\aSclCt}{\vartheta}
\newcommand{\aVecCt}{\aVecC}
\newcommand{\aSclD}{\rho}
\newcommand{\aVecD}{\bm{\aSclD}}
\newcommand{\aSclE}{\tau}
\newcommand{\aSclF}{\nu}
\newcommand{\aVecF}{\bm{\aSclF}}
\newcommand{\aSclG}{\upsilon}
\newcommand{\aVecG}{\bm{\aSclG}}
\newcommand{\aMatG}{\Upsilon}
\newcommand{\aSclH}{\aSclE}
\newcommand{\aVecH}{\bm{\aSclH}}
\newcommand{\aVecGrI}{\bm{\upsilon}}

\newcommand{\tempcumul}{\kappa}
\newcommand{\cumul}[2]{\tempcumul_{#1,#2}}
\newcommand{\realcumul}{\tempcumul}
\newcommand{\seccumul}{\cumul{2}{0}}
\newcommand{\indexset}{\mathcal{I}}
\newcommand{\emptyindex}{\varnothing}
\newcommand{\diLambda}{\mathcal{M}}
\newcommand{\stpointsnbh}{\Omega_n}
\newcommand{\idom}{\mathcal{D}}
\newcommand{\Vanddet}{\triangle}
\newcommand{\herm}{\mathcal{H}}
\newcommand{\USp}{\mathrm{USp}}
\newcommand{\scoeff}{\mathsf{c}}
\newcommand{\vol}{\mathrm{vol}}
\newcommand{\skB}{\mathcal{B}}
\newcommand{\Hess}{\mathsf{H}}
\newcommand{\Fperm}{\mathcal{F}}

\newcommand{\partder}[2]{\frac{\partial #1}{\partial #2}}
\newcommand{\der}[2]{\frac{d #1}{d #2}}

\newcommand{\abs}[1]{\left\lvert#1\right\rvert}
\newcommand{\abssized}[2][ ]{#1\lvert#2#1\rvert}
\newcommand{\norm}[1]{\left\lVert#1\right\rVert}
\newcommand{\normsized}[2][ ]{#1\lVert#2#1\rVert}

\newcommand{\tr}{\mathop{\mathrm{tr}}}
\newcommand{\Pf}{\mathop{\mathrm{Pf}}}
\newcommand{\E}{\operatorname{\mathbf{E}}}
\newcommand{\diag}{\mathop{\mathrm{diag}}}



\begin{abstract}
We study asymptotic behaviour of the correlation functions of
bipartite sparse weighted random $N\times N$ matrices. 
It is shown that the main
term of the correlation function of $k$-th and $m$-th moments of the
integrated density of states is  $N^{-1}n_{k,m}$. The closed
system of recurrent relations for coefficients
$\{n_{k,m}\}_{k,m=1}^\infty$ was obtained.
\key{bipartite sparse random graph, correlator of moments, asymptotics, main term,  system of recurrence equations.}

\msc{60B20, 15B52.}
\end{abstract}


\section{Introduction}

In the last few years interest in the spectral properties of ensembles of  sparse random matrices has sharply increased.
It is expected that the spectral properties of sparse random matrices will differ from the properties of the ensembles of most matrices with independent elements
 (see \cite{W}, as well as the survey works \cite{KKPS}, \cite{Pa:00} and the literature cited therein).

 Interesting results for sparse random matrices have been obtained in a series of physical works \cite{RB:88}, \cite{RD:90}, \cite{MF:91},
  \cite{FM:96}. In particular, the equation for the Laplace transform of
  limiting integrated state density is derived, studied
  "density-density correlator" is studied and it is shown that there exists some
  critical point $p_c> 1$, in the vicinity of which occurs
  phase transition in $p$: for $p<p_c$ all eigenvectors are localized,
  whereas for $p>p_c$ delocalized eigenvectors appear.
Unfortunately all these results were obtained by replica or supersymmetry methods, and therefore need mathematically
  correct justification.

In a series of mathematical papers \cite{BG1, BG2, B} it is proved
  the existence of a limit for $N\to\infty$ averaged moments
  integrated state density in the simplest case, when the matrix elements are equal to $0$
 with probability $1-p/N$ and $1$ with probability $p/N$. It is shown, that
 limiting moments satisfy the Carleman condition, thereby
 the existence of a limit of integrated state density for the  ensemble of sparse random matrices is proved.
In the papers \cite{KV}, \cite{KSV} similar results were obtained for wider class of  sparse random matrix ensemble. In papers \cite{PJ}, \cite{CS}  delocalization and an existence of the absolute continious part of the limiting spectra at 0 were studied.
 In the  paper \cite{V0} the behavior of the asymptotics of the correlator of moments
 as $ N \to\infty $ was studied.

In the papers \cite{V}, \cite{V1} similar results were obtained for the  bipartite  sparse random matrix ensemble.
In this  paper we study the behavior of the asymptotics of the correlator of moments
 as $N\to\infty $ for the  bipartite  sparse random matrix ensemble. The speed of approaching to zero  and the value of the  main term are very important in physical
applications. Therefore an extensive literature is devoted to similar studies for various ensembles of random matrices (see.
for example, the works \cite{KKP}, \cite{APS2} and quoted in them literature).


\section{Main results}


    We can introduce the randomly weighted
adjacency matrix of random bipartite graphs.
Let $\Xi=\{a_{ij} ,\; i \!\leq\! j,\;i,j\! \in \!{\mathbb N}\}$ be
the  set of
jointly independent identically distributed (i.i.d.) random
variables determined
  on the same probability space and possessing the moments
\begin{equation}\label{const_all_mom}
  {\mathbb E}a^{k}_{ij}\!=\!X_{k}\!<\!\infty \qquad
  \forall \; i,j,k\in {\mathbb N},
\end{equation}
where ${\mathbb E}$ denotes the mathematical expectation corresponding to
$\Xi$.
  We set $a_{ji}\!=\! a_{ij}$ for $i\!\leq\! j \;$.

Given $ 0\!<p\!\leq \!N$, let us define the family
  $B^{(p)}_N\!=\!\{b^{(N,p)}_{ij},
\; i\!\leq\! j,\; i,j\in \overline{1,N}\}$ of jointly independent
random variables
\begin{equation}
b^{(N,p)}_{ij}\!=\! \left\{ \begin{array}{ll} p^{-1/2},&
\textrm{with} \ \textrm{probability } \ p/N ,
\\0,& \textrm{with} \ \textrm{probability} \ 1-p/N ,\\ \end{array}
\right.
\end{equation}
We determine $b^{(N,p)}_{ji}= b^{(N,p)}_{ij}$ and assume that $ B^{(p)}_N$
is independent from $\Xi$.

 Let $\alpha \in (0,1)$, denote by $I_1^{(N,\alpha)}=\overline{1,\lfloor\alpha \cdot N\rfloor}$, $I_2^{(N,\alpha)}=\overline{\lfloor\alpha \cdot N\rfloor+1,N}$, where $\lfloor \cdot \rfloor$ is a floor function. Now one can consider the real symmetric $N\times N$ matrix
$A^{(N,p,\alpha)}(\omega)$:
\begin{equation}\label{dilute}
A^{(N,p,\alpha)}_{ij}\!=\! a_{ij}\cdot b_{ij}^{(N,p)}\cdot \xi^{(N,\alpha)}_{ij},
\end{equation}
\begin{equation}\label{delta}
\textrm{where}\ \xi^{(N,\alpha)}_{i,j}\!=\!\left\{ \begin{array}{ll} 1,&
\textrm{if} \  (i \in I_1^{(N,\alpha)} \wedge j\in I_2^{(N,\alpha)} ) \vee (i \in I_2^{(N,\alpha)} \wedge j\in I_1^{(N,\alpha)} )  ,
\\0,& \textrm{otherwise} \\ \end{array}
\right.
\end{equation}
that has $N$ real eigenvalues
$\lambda^{(N,p,\alpha)}_1\!\leq\!\lambda^{(N,p,\alpha)}_2 \!\leq\!\ \ldots
\!\leq\!\ \lambda^{(N,p,\alpha)}_N$.

The normalized eigenvalue counting function (or integrated density
of states (IDS)) of  $A^{(N,p,\alpha)}$ is determined
by the formula
$$
\sigma\left({\lambda; A^{(N,p,\alpha)}}\right)\!=\!\frac{\sharp
\{j:\lambda^{(N,p,\alpha)}_j\!<\!\lambda\}}{N}.
$$
The following denotations are used:
$$
{\cal{M}}^{(N,p,\alpha)}_k\!=\! \int \lambda^k d \sigma\left({\lambda;
A^{(N,p,\alpha)}}\right), \ M^{(N,p,\alpha)}_k=\mathbb{E}{\cal{M}}^{(N,p,\alpha)}_k,
$$
$$
C^{(N,p,\alpha)}_{k,m}= \mathbb{E}\left\{{\cal{M}}^{(N,p,\alpha)}_{k}\cdot {\cal{M}}^{(N,p,\alpha)}_{m} \right\}-\mathbb{E}\left\{{\cal{M}}^{(N,p,\alpha)}_{k} \right\}\cdot \mathbb{E}\left\{{\cal{M}}^{(N,p,\alpha)}_{m} \right\} .
$$

\begin{theorem}
\label{thm:1}
 Main asymptotic coefficients of correlators $n^{(p,\alpha)}_{k,m}$ 
\begin{equation}\label{main_prop}
\lim_{N\to\infty}N\cdot C^{(N,p,\alpha)}_{k,m}\!=\! \left\{ \begin{array}{ll} n^{(p,\alpha)}_{k/2,m/2},&
\textrm{if} \ k \ \textrm{and } \ m \  \textrm{are even }
\\0,&  \textrm{otherwise} \\ \end{array}
\right.
\end{equation}
  can be obtained by the system of recurrent relations   $\ref{rec}-\ref{rec_=c}$, $\ref{rec_g=c}-\ref{rec_rdn=c}$
 with the initial conditions $\ref{ini_1}-\ref{ini_9}$.
\end{theorem}

\section{Proof of Theorem 1}
\subsection{Correlators and double bipartite walks}

 Let us transform correlator $C^{(N,p,\alpha)}_{k,m}$ to convenient form for the limiting transition.

$$
C^{(N,p,\alpha)}_{k,m}\!=\! {\mathbb E}\left\{ {\cal
M}^{(N,p,\alpha)}_k\cdot{\cal
M}^{(N,p,\alpha)}_m\right\}-{\mathbb E}\left\{ {\cal
M}^{(N,p,\alpha)}_k\right\}\cdot{\mathbb E}\left\{ {\cal
M}^{(N,p,\alpha)}_m\right\}\!=
$$
$$
=\!\frac{1}{N^2}\left({\mathbb E}\left\{\mathrm{Tr}
[A^{(N,p,\alpha)}]^k \cdot \mathrm{Tr}
[A^{(N,p,\alpha)}]^m\right\}-{\mathbb E}\left\{\mathrm{Tr}
[A^{(N,p,\alpha)}]^k \right\}\cdot{\mathbb E}\left\{\mathrm{Tr}
[A^{(N,p,\alpha)}]^m \right\}\right)\!=
$$
$$
\!=\frac{1}{N^2} \sum^{N}_{i_1,\ldots,i_k=1}
 \sum^{N}_{j_1,\ldots,j_m=1}
 \left({\mathbb E}\left\{A^{(N,p,\alpha)}_{i_1,i_2}
 A^{(N,p,\alpha)}_{i_2,i_3} \ldots
 A^{(N,p,\alpha)}_{i_{k},i_1}\cdot\right.\right.
$$
$$
 \left.\cdot A^{(N,p,\alpha)}_{j_1,j_2}
 A^{(N,p,\alpha)}_{j_2,j_3} \ldots
 A^{(N,p,\alpha)}_{j_{m},j_1}
 \right\}-{\mathbb E}\left\{ A^{(N,p,\alpha)}_{i_1,i_2}
 A^{(N,p,\alpha)}_{i_2,i_3} \ldots
 A^{(N,p,\alpha)}_{i_{k},i_1}\right\} \cdot
$$
$$
  \left.\cdot{\mathbb E}\left\{ A^{(N,p,\alpha)}_{j_1,j_2}
 A^{(N,p,\alpha)}_{j_2,j_3} \ldots
 A^{(N,p,\alpha)}_{j_{m},j_1} \right\}\right)\!=\!\frac{1}{N^2}
\sum^{N}_{i_1=1} \sum^{N}_{i_2=1} \ldots
  \sum^{N}_{i_{k}=1}
 \sum^{N}_{j_1=1} \sum^{N}_{j_2=1} \ldots
  \sum^{N}_{j_{m}=1}
$$
$$
   \bigg( {\mathbb E}\left\{
 a_{i_1,i_2}\cdot
 a_{i_2,i_3}\cdot \ldots \cdot a_{i_{k},i_1} \cdot
 a_{j_1,j_2}\cdot
 a_{j_2,j_3}\cdot \ldots \cdot a_{j_{m},j_1}\right\} \cdot
 $$
 $$
 \cdot {\mathbb E}\left\{ b^{(N,p)}_{i_1,i_2}
 b^{(N,p)}_{i_2,i_3}\cdot \ldots \cdot b^{(N,p)}_{i_{k},i_1}\cdot
  b^{(N,p)}_{j_1,j_2}
 b^{(N,p}_{j_2,j_3}\cdot \ldots \cdot b^{(N,p)}_{j_{m},j_1}
 \right\}-
 $$
 $$
 {\mathbb E}\left\{ a_{i_1,i_2}
 a_{i_2,i_3}\cdot \ldots \cdot a_{i_{k},i_1}\right\}\cdot{\mathbb E}
  \left\{  a_{j_1,j_2}\cdot
 a_{j_2,j_3}\cdot \ldots \cdot a_{j_{m},j_1}\right\}\cdot
 $$
$$
 \cdot{\mathbb E}
 \left\{b^{(N,p)}_{i_1,i_2}
 b^{(N,p)}_{i_2,i_3}\cdot \ldots \cdot b^{(N,p)}_{i_{k},i_1} \right\} \cdot
 {\mathbb E}\left\{ b^{(N,p)}_{j_1,j_2}
 b^{(N,p)}_{j_2,j_3}\cdot \ldots \cdot b^{(N,p)}_{j_{m},j_1}
 \right\}\bigg)
$$
\begin{equation}\label{eq:cor1}
\cdot  \xi^{(N,\alpha)}_{i_1,i_2} \cdot
  \xi^{(N,\alpha)}_{i_2,i_3} \cdot \ldots \cdot\xi^{(N,\alpha)}_{i_{k},i_1}\cdot  \xi^{(N,\alpha)}_{j_1,j_2} \cdot
  \xi^{(N,\alpha)}_{j_2,j_3} \cdot \ldots \cdot\xi^{(N,\alpha)}_{j_{m},j_1}
\end{equation}

 \qquad Let $W^{(N,\alpha)}_{k}$ be a set of closed bipartite walks of $k$ steps over
the sets $I_1^{(N,\alpha)}=\overline{1,\lfloor\alpha\cdot N\rfloor}$ and $I_2^{(N,\alpha)}=\overline{\lfloor\alpha\cdot N\rfloor+1,N}$:
$W^{(N,\alpha)}_{k}\!=\!{}^{(1)}W^{(N,\alpha)}_{k}\cup {}^{(2)}W^{(N,\alpha)}_{k}$, where
$$
 {}^{(1)}W^{(N,\alpha)}_{k}\!=\!\{w\!=\!(w_1,w_2,\cdots,w_k,w_{k+1}=w_1):
\forall i \!\in\! \overline{1,k+1} \;\: w_i\!\in\!
I_{2-(i\, \mathrm{mod} \,  2)}^{(N,\alpha)}\} ,
$$
$$
 {}^{(2)}W^{(N,\alpha)}_{k}\!=\!\{w\!=\!(w_1,w_2,\cdots,w_k,w_{k+1}=w_1):
\forall i \!\in\! \overline{1,k+1} \;\: w_i\!\in\!
I_{1+(i\, \mathrm{mod} \,  2)}^{(N,\alpha)}\}.
$$
Here (a mod m) denotes the residue of a modulo m. Thus, for a walk $w$ from $W^{(N,\alpha)}_{k}$, either all odd elements are from $I_1^{(N,\alpha)}$, and even ones are from $I_2^{(N,\alpha)}$, or vice versa, all odd elements are from $I_2^{(N,\alpha)}$, and all even ones are from $I_1^{(N,\alpha)}$.  In the first case, $w$ is from ${}^{(1)}W^{(N,\alpha)}_{k}$, and in the second case, $w$ is from ${}^{(2)}W^{(N,\alpha)}_{k}$. Here are some examples. $(2,4,1,5,3,4,2,6,2)\in {}^{(1)}W^{(6,1/2)}_{8}$. $(5,1,4,2,6,3,6,1,5)\in {}^{(2)}W^{(6,1/2)}_{8}$. $(2,4,1,5,3,4,2,6,1)\not\in {}^{(1)}W^{(6,1/2)}_{8}$ (nonclosed). $(2,4,1,5,3,4,2,3,2)\in {}^{(1)}W^{(6,1/2)}_{8}$ (nonbipartite).
For $w\!\in\!W^{(N,\alpha)}_k$ let us denote
  $a(w)\!=\!\prod_{i=1}^{k} a_{w_i,w_{i+1}}$,
 $b^{(N,p)}(w)\!=\!\prod_{i=1}^{k} b^{(N,p)}_{w_i,w_{i+1}}$ .

  Let  $\mathfrak{D}^{(N,\alpha)}_{k,m}\stackrel{\rm def}{\equiv}W^{(N,\alpha)}_{k} \times
W^{(N,\alpha)}_{m}$ be a set of double bipartite walks of $k$ and $m$ steps over
the sets $I_1^{(N,\alpha)}$ and $I_2^{(N,\alpha)}$.
For
$d\!=\!(w^{(1)},w^{(2)}) \!\in\!\mathfrak{D}^{(N)}_{k,m}$ let us denote
$$
a(d)\!=\!a(w^{(1)})\cdot a(w^{(2)}),\;\;
b^{(N,p)}(d)\!=\!b^{(N,p)}(w^{(1)})\cdot b^{(N,p)}(w^{(2)}).
$$
 Then we can rewrite equality
 (\ref{eq:cor1}) in the following way :
$$
C^{(N,p,\alpha)}_{k,m}\!=\!\frac{1}{N^2} \sum_{d=(w^{(1)},w^{(2)})\in
W^{(N,\alpha)}_{k,m}} \left\{ {\mathbb E} a(d) \cdot {\mathbb E}
b^{(N,p)}(d) -\right.
$$
\begin{equation}\label{eq:cor2}
 \left.-{\mathbb E} a(w^{(1)}) \cdot {\mathbb E}
b^{(N,p)}(w^{(1)}) \cdot {\mathbb E} a(w^{(2)}) \cdot {\mathbb E}
b^{(N,p)}(w^{(2)})\right\}.
\end{equation}
For $w\!\in\! W^{(N,\alpha)}_k$
  and
$f,g \!\in\! \overline{1,N}\ $ denote by $n_w(f,g)$ the number
of steps $f\to g$ and $g \to f$;
$$
n_w(f,g)=\#\{i \!\in\! \overline{1,k}:\; (w_i\!=\!f \ \wedge \
w_{i+1}\! =\!g)\vee (w_i\!=\!g \ \wedge \ w_{i+1}\!=\!f)\}.
$$

 For $w\!=\!(w^{(1)},w^{(2)}) \!\in\!\mathfrak{D}^{(N,\alpha)}_{k,m}$
let us introduce similar denotation
$$
n_{d}(f,g)=n_{w^{(1)}}(f,g)+n_{w^{(2)}}(f,g).
$$
 Then for all $\ w \!\in\! W^{(N,\alpha)}_k$ and all  $\ d \!\in\!
\mathfrak{D}^{(N,\alpha)}_{k,m}$ we have
$$
{\mathbb E}a(w)\!=\! \prod_{f=1}^{N} \prod_{g=f}^{N} V_{n_w(f,g)}
\qquad {\mathbb E}a(d)\!=\! \prod_{f=1}^{N} \prod_{g=f}^{N}
V_{n_{d}(f,g)}.
$$
 Given $w\!\in\! W^{(N)}_k$,
let us define the sets $V_w=\cup_{i=1}^{k}\{w_i\}$ and
$E_w=\cup_{i=1}^{k}\{(w_i,w_{i+1})\},$ where $(w_i,w_{i+1})$ is a
non-ordered pair. It is easy to see that $G_w\!=\!(V_w,E_w)$ is a
simple connected non-oriented bipartite graph and the walk $w$ covers the graph
$G_w$. Let us call $G_w$ the skeleton of walk $w$. We denote by
$n_w(e)$ the number of passages of the edge $e$ by the walk $w$ in
direct and inverse directions. For
$(w_j,w_{j+1})\!=\!e_j\!\in\!E_w$ let us denote
$a_{e_j}\!=\!a_{w_j,w_{j+1}}\!=\!a_{w_{j+1},w_j}$. Then we obtain
$$
  {\mathbb
E}a(w)\!=\!\prod_{e\in E_w} {\mathbb E}a^{n_w(e)}_e\!=\! \prod_{e\in
E_w} V_{n_w(e)}.
  $$
Similarly we can write

$$
{\mathbb E}b^{(N,p)}(w)\!=\!\prod_{e\in E_w} {\mathbb E}
\left([b^{(N,p)}_e]^{n_w(e)}\right)\!=\! \prod_{e\in
E_w} \frac{1}{N\cdot p^{n_w(e)/2-1}}.
$$
 Let's introduce similar definitions for double bipartite walk
$d=(w^{(1)},w^{(2)})\!\in\!\mathfrak{D}^{(N)}_{k,m}$.
$V_{d}=V_{w^{(1)}}\cup V_{w^{(2)}}$, $E_{d}=E_{w^{(1)}}\cup
E_{w^{(2)}}$, $G_{d}\!=\!(V_{d},E_{d})$ the following equations hold
$${\mathbb E}a(d)\!=\! \prod_{e\in E_{d}} V_{n_{d}(e)} \qquad
{\mathbb E}b^{(N,p)}(d)\!=\! \prod_{e\in E_{d}} \frac{1}{N\cdot
p^{n_{d}(e)/2-1}}.$$
 Then, we can rewrite (\ref{eq:cor2}) in the form
$$
C^{(N,p,\alpha)}_{k,m}\!=\! \frac{1}{N^2} \sum_{d=(w^{(1)},w^{(2)})\in
W^{(N,\alpha)}_{k,m}} \left\{ \prod_{e\in E_{d}}{\mathbb E}
a_e^{n_{d}(e)} \cdot {\mathbb E} \left[b^{(N,p)}_e \right]^{n_{d}(e)} -\right.
$$
$$ -\left.\prod_{e\in
E_{w^{(1)}}}{\mathbb E} a_e^{n_{w^{(1)}}(e)} \cdot {\mathbb E}
\left[ b^{(N,p)}_e\right]^{n_{w^{(1)}}(e)}  \cdot \prod_{e\in E_{w^{(2)}}}{\mathbb E}
a_e^{n_{w^{(2)}}(e)} \cdot {\mathbb E} \left[ b^{(N,p)}_e\right]^{n_{w^{(2)}}(e)}\right\}\!=
$$
$$
=\! \frac{1}{N^2} \sum_{d=(w^{(1)},w^{(2)})\in
W^{(N,\alpha)}_{k,m}} \left\{N^{-|E_{d}|}\cdot \prod_{e\in E_{d}}\frac{V_{n_{d}(e)}}{p^{{n_{d}(e)}/2-1}}
 -\right.
$$
$$
\left. N^{-|E_{w^{(1)}}|-|E_{w^{(2)}}|}\cdot
\prod_{e\in E_{w^{(1)}}}\frac{V_{n_{w^{(1)}}(e)}}{p^{{n_{w^{(1)}}(e)}/2-1}}
\prod_{e\in E_{w^{(2)}}}\frac{V_{n_{w^{(2)}}(e)}}{p^{{n_{w^{(2)}}(e)}/2-1}}\right\}\!=
$$
$$
=\! \frac{1}{N^2} \sum_{d=(w^{(1)},w^{(2)})\in W^{(N,\alpha)}_{k,m}}
\left\{\frac{1}{N^{|E_{d}|}\cdot p^{(k+m)/2-|E_{d}|}}\cdot
\prod_{e\in E_{d}}V_{n_{d}(e)} -\right.
$$
\begin{equation}\label{eq:walks}
\left.\frac{1}{ N^{|E_{w^{(1)}}|+|E_{w^{(2)}}|}\cdot
p^{(k+m)/2-|E_{w^{(1)}}|-|E_{w^{(2)}}|}}\cdot \prod_{e\in
E_{w^{(1)}}}V_{n_{w^{(1)}}(e)} \prod_{e\in
E_{w^{(2)}}}V_{n_{w^{(2)}}(e)}\right\}\ \!=\!\sum_{d\in
W^{(N,\alpha)}_{k,m}}\theta(d),
\end{equation}
where $\theta(d)$ is the contribution of the double bipartite walk $d$ to the
mathematical
  expectation of the corresponding correlator.The last expression is not very convenient for the  limiting transition. Moreover,
the latter formula shows that the contribution of a double bipartite walk $d$ depends only  on sets
\begin{equation} \label{conditions}
\cup_{e\in E_{d}}\{n_{d}(e)\},\ \cup_{e\in E_{w^{(1)}}}\{n_{w^{(1)}}(e)\},\
\cup_{e\in E_{w^{(2)}}}\{n_{w^{(2)}}(e)\}.
\end{equation}
Therefore,
it is natural to introduce an equivalence relation on $\mathfrak{D}^{(N,\alpha)}_{k,m}$. Double bipartite walks
$ d=(w^{(1)},w^{(2)}), u=(u^{(1)},u^{(2)})  \in W^{(N,\alpha)}_{k,m}$ are equivalent $d\sim u$ if and only if
there exists a partition preserving bijection $\phi$ between the sets of
vertices $V_{d}$ and  $V_{u}$ such that for
$$
 \ d \sim u \Leftrightarrow \exists \phi:  V_{d}\stackrel{bij}{\to}
 V_{u}: \quad  \phi(V_{d}\cap I_1^{(N,\alpha)}) =  V_{u}\cap I_1^{(N,\alpha)},
  u\!=\!\phi(d)
$$
 Let us denote by $[d]$ the class of equivalence of double bipartite walk $d$ and by
$\mathfrak{C}^{(N,\alpha)}_{k,m}$ the set of such classes for all $d \in \mathfrak{D}^{(N,\alpha)}_{k,m}$. It is obvious that if two
walks $d$ and $u$ are equivalent then their
contributions are equal:
$$
  d\sim u\;\Longrightarrow
\theta(d)\!=\!\theta(u).
$$
Cardinality of the class of equivalence $[d]$ is equal the number
of all mappings $\phi:V_d \to  \overline{1,N}$ such that $\phi(V_{1,d}) \subset  I_1^{N,\alpha}$ and $\phi(V_{2,d}) \subset  I_2^{N,\alpha}$ (where $V_{1,d}=V_{d}\cap I_1^{(N,\alpha)}$ and  $V_{2,d}=V_{d}\cap I_2^{(N,\alpha)}$). Therefore, it  is equal to the number $\lfloor\alpha\cdot N\rfloor
\cdot (\lfloor\alpha\cdot N\rfloor-1) \cdot \ldots \cdot (\lfloor\alpha\cdot N\rfloor-|V_{1,d}|+1)\cdot (\lceil(1-\alpha)\cdot N\rceil)
\cdot (\lceil(1-\alpha)\cdot N\rceil-1)\cdot \ldots \cdot (\lceil(1-\alpha)\cdot N\rceil-|V_{2,d}|+1)$. Then we can rewrite
(\ref{eq:walks}) in the following form
$$
 C^{(N,\alpha)}_{k,m}=\!\frac{1}{N^2} \sum_{[d]\in \mathfrak{C}^{(N,\alpha)}_{k,m}}
\left\{\frac{\lfloor\alpha_1\cdot N\rfloor
\cdot (\lfloor\alpha_1\cdot N\rfloor-1) \cdot \ldots \cdot (\lfloor\alpha_1\cdot N\rfloor-|V_{1,d}|+1)}{N^{|E_{d}|}\cdot p^{(k+m)/2-|E_{d}|}} \right.
$$
$$
\cdot (\lceil\alpha_2\cdot N\rceil)
\cdot (\lceil\alpha_2\cdot N\rceil-1)\cdot \ldots \cdot (\lceil\alpha_2\cdot N\rceil-|V_{2,d}|+1)\cdot\left(
\prod_{e\in E_{d}}V_{n_{d}(e)} -\right. 
$$
\begin{equation}\label{eq:class}
\left.\left.\frac{p^{|E_{w^{(1)}}|+|E_{w^{(2)}}|-|E_{d}|}}
{N^{|E_{w^{(1)}}|+|E_{w^{(2)}}|-|E_{d}|}} \cdot \prod_{e\in
E_{w^{(1)}}}V_{n_{w^{(1)}}(e)} \prod_{e\in
E_{w^{(2)}}}V_{n_{w^{(2)}}(e)}\right)\right\}\!=\!\sum_{[d]\in
\mathfrak{C}^{(N,\alpha)}_{k,m}} \theta([d]),
\end{equation}
where $\alpha_1=\alpha$ and $\alpha_2=1-\alpha$.

But the transition to the limit $ N\to\infty $ in the last formula is hindered by the dependence of $\mathfrak{C}^{(N,\alpha)}_{k, m}$ on $N$.
In order to solve this problem, and at the same time
for better understanding of $ \mathfrak{C}^{(N,\alpha)}_{k,m}$ we introduce the notion of minimal double bipartite
walks.
\subsection{Minimal and essential walks}

It is convenient to deal with $\widetilde{\mathfrak{D}}^{(N,\alpha)}_{k,m}$ instead of $ \mathfrak{D}^{(N,\alpha)}_{k,m}$, where $\widetilde{\mathfrak{D}}^{(N,\alpha)}_{k,m}$
 is a set of double  bipartite closed walks over the sets $I_1^{(N,\alpha)}$ and $\widetilde{I}_2^{(N,\alpha)}=\left\{\widetilde{1},\widetilde{2}\ldots,
 \widetilde{\lceil N\cdot(1-\alpha)\rceil}\right\}$.
We just renamed the vertices of the second component. Let us consider $\widetilde{\mathfrak{C}}^{(N,\alpha)}_{k,m}$, the set of equivalence  classes of $\widetilde{\mathfrak{D}}^{(N,\alpha)}_{k,m}$. As a  representative of the equivalence  class $[d]\in \widetilde{\mathfrak{C}}^{(N,\alpha)}_{k,m}$ , we can take a minimal double walk.
\begin{definition}
   A double bipartite closed walk  $d\in \widetilde{\mathfrak{C}}^{(N,\alpha)}_{k,m}$ is called minimal if and only if at each stage of the passage a new vertex is the minimum element among the unused vertices of the corresponding component. In this case, we apply the following procedure for passing a double walk: first we pass the first walk, then we jump over to the initial vertex of the second walk and then we pass it.
\end{definition}
Let us denote the set of all minimal walks of $\widetilde{\mathfrak{D}}^{(N,\alpha)}_{k,m}$ by
$\mathfrak{M}^{(N,\alpha)}_{k,m}$.
\vskip 0.5cm

\noindent {\bf Example 1.} The double walk  $\left((1,\widetilde{1},1,\widetilde{2},1),
(\widetilde{3},2,\widetilde{3},1,\widetilde{1},1,\widetilde{3})\right)$ is the   minimal one.

 Then (\ref{eq:class} ) can be written as
 $$
 C^{(N,\alpha)}_{k,m}=\!\frac{1}{N^2} \sum_{d\in \mathfrak{M}^{(N,\alpha)}_{k,m}}
\left\{\frac{\lfloor\alpha_1\cdot N\rfloor
\cdot (\lfloor\alpha_1\cdot N\rfloor-1) \cdot \ldots \cdot (\lfloor\alpha_1\cdot N\rfloor-|V_{1,d}|+1)}{N^{|E_{d}|}\cdot p^{(k+m)/2-|E_{d}|}}\right.
$$
$$
\cdot (\lceil\alpha_2\cdot N\rceil)
\cdot (\lceil\alpha_2\cdot N\rceil-1)\cdot \ldots \cdot (\lceil\alpha_2\cdot N\rceil-|V_{2,d}|+1)\cdot\left(
\prod_{e\in E_{d}}V_{n_{d}(e)} -\right.
$$
\begin{equation}\label{eq:min}
\left.\left.\frac{p^{|E_{w^{(1)}}|+|E_{w^{(2)}}|-|E_{d}|}}
{N^{|E_{w^{(1)}}|+|E_{w^{(2)}}|-|E_{d}|}} \cdot \prod_{e\in
E_{w^{(1)}}}V_{n_{w^{(1)}}(e)} \prod_{e\in
E_{w^{(2)}}}V_{n_{w^{(2)}}(e)}\right)\right\}\!=\!\sum_{d\in
\mathfrak{M}^{(N,\alpha)}_{k,m}} \theta([d]),
\end{equation}

  Each double walk $d \in \mathfrak{M}^{(N,\alpha)}_{k,m}$ has at most $k+m$ vertices.
    Hence, $\mathfrak{M}^{(1,\alpha)}_{k,m}
\subset \mathfrak{M}^{(2,\alpha)}_{k,m} \subset \ldots \subset \mathfrak{M}^{(\lceil(k+m)\cdot\min(\alpha_1,\alpha_2)^{-1})\rceil,\alpha)}_{k,m}  = \mathfrak{M}^{(\lceil(k+m)\cdot\min(\alpha_1,\alpha_2)^{-1})\rceil+1,\alpha)}_{k,m}= \ldots$. It is natural to denote
  $\mathfrak{M}^{(\alpha)}_{k,m}=\mathfrak{M}^{(\lceil(k+m)\cdot\min(\alpha_1,\alpha_2)^{-1})\rceil,\alpha)}_{k,m}$.
  Let us denote the number of common edges of $G_{w{(1)}}$ and $G_{w{(2)}}$ by
$c(d)=|E_{w^{(1)}}|+|E_{w^{(2)}}|-|E_{d}|$.
Then the following equality for the main asymptotic coefficient of the correlator holds
$$
 n^{(p,\alpha)}_{k/2,m/2} \!=\! \lim_{N \to \infty} N\cdot C^{(N,p,\alpha)}_{k,m} \!=\!
\sum_{w\in \mathfrak{M}^{(\alpha)}_{k,m}} \lim_{N \to \infty}
\left[\frac{N^{|V_{d}|-|E_{d}|-1}}
 {p^{(k+m)/2-|E_{d}|}}\cdot \alpha_1^{|V_{1,d}|}\cdot \alpha_2^{|V_{2,d}|}\cdot \right.
$$
\begin{equation}\label{eq:min_lim}
\left. \left(
\prod_{e\in E_{d}}V_{n_{d}(e)} -  \frac{p^{c(d)}} {N^{c(d)}} \cdot
\prod_{e\in E_{w^{(1)}}}V_{n_{w^{(1)}}(e)} \prod_{e\in
E_{w^{(2)}}}V_{n_{w^{(2)}}(e)}\right) \right] .
\end{equation}

  $\mathfrak{M}^{(\alpha)}_{k,m}$ is a finite set.
Not all minimal walks make a non-zero contribution to the main asymptotic coefficient of the correlator.
  $G_{d}$ has at most 2 connected components, because  $G_{w{(1)}}$ and $G_{w{(2)}}$ are connected graphs.
But if graph  $G_{d}$ has exactly 2 connected components, then  $V_{w^{(1)}}\cap V_{w^{(2)}}\!=\!\varnothing
\Rightarrow E_{w^{(1)}}\cap E_{w^{(2)}}\!=\!\varnothing
\Rightarrow c(d)=0 \Rightarrow$
$$\Rightarrow \left( \prod_{e\in
E_{d}}V_{n_{d}(e)} - \frac{p^{c(d)}} {N^{c(d)}} \cdot
\prod_{e\in E_{w^{(1)}}}V_{n_{w^{(1)}}(e)} \prod_{e\in
E_{w^{(2)}}}V_{n_{w^{(2)}}(e)}\right)\!=\! 0.
$$
 Consequently, such minimal double bipartite walks make zero contribution to
$ n^{(p,\alpha)}_{k/2,m/2}$.

This means that only minimal double walks with a connected skeleton $ G_ {d} $ can make a nonzero contribution. For any connected graph
$ G_{d} $ the inequality $ | V_{d} | - | E_{d} | -1 \leq 0 $ holds, and the equality holds
if and only if only if $ G_{d} $
is a tree. There are two cases: $ E_{w^{(1)}}\cap
E_{w^{(2)}}\!=\!\varnothing \Rightarrow c(d)=0$ and $c(d)>0$.
In the first case, the contribution is 0 (see above), and in the second, it is
$\alpha_1^{|V_{1,d}|}\cdot \alpha_2^{|V_{2,d}|}\cdot\prod_{e\in E_{d}}\frac{V_{n_{d}(e)}}{p^{n_{d}(e)/2-1}} .$
\begin{definition}
We call essential those minimal double bipartite walks whose contribution to the main asymptotic coefficient of the corresponding correlator is not equal to 0.
\end{definition}
Denote the set of essential double walks by $\mathfrak{S}_{k,m}$. $\mathfrak{S}_{k,m}\!=\!\{d\in \mathfrak{M}_{k,m}:G_{d}\textrm{ is  a tree }
\wedge c(d)>0 \}.$ These are all minimal double bipartite walks, whose graph is a tree and the graphs of the first and second
 walks have at least one common edge. Now the formula \ref{eq:min_lim}
can be written like this
\begin{equation}\label{eq:ess}
n^{(p,\alpha)}_{k/2,m/2}\!=\!\sum_{d\in  \mathfrak{S}_{k,m}}\theta(d),
\quad\textrm{where  }\quad   \theta(d)\!=\!\alpha_1^{|V_{1,d}|}\cdot \alpha_2^{|V_{2,d}|}\cdot\prod_{e\in
E_{d}}\frac{V_{n_{d}(e)}}{p^{n_{d}(e)/2-1}}.
\end{equation}
So we proved the first part of the theorem, the existence of $n^{(p)}_{k/2,m/2}$. It remains to derive a system of recurrence relations for $n^{(p)}_{k/2, m/2}$.
From the definition it is clear that the weight $\theta$ of a essential double bipartite walk is multiplicative along the edges  of
$G_{d}$.

Clearly, that $\mathfrak{S}_{k,m}=\emptyset$ if $k$ is odd  or $m$ is odd. Indeed, from  definition of essential double walk $d$ it follows that $G_d$ is a tree. Hence, $G_{w^{(1)}}$ and $G_{w^{(2)}}$ are trees. Each edge of any tree is a bridge. Thus, since $w^{(1)}$ and $w^{(2)}$ are closed walks, then their lengths are even numbers. So  $n^{(p)}_{k/2,m/2}=0$ for such $k$ and $m$.
\subsection{First edge decomposition of essential walks}
  The idea of derivation of the recurrent system is the same as that of Wigner (\cite{W}), but its implementation is more complicated.
Remove from the graph $G_{d}$ the first edge $(r,v)$ of the first walk $w^{(1)}$. Since $G_{d}$ is a tree, the graph splits into two pieces: an upper graph $G_u$ , which
contains vertex $v$, and right graph $G_r$, which contains vertex $r$. Then bipartite walk $w^{(1)}$ (resp. $w^{(2)}$) is divided into an upper bipartite walk
$w^{(1,u)}$ (resp. $w^{(2,u)}$) on $G_u$ and a right bipartite walk $w^{(1,r)}$ (resp. $w^{(2,r)}$) on $G_r$. Similarly, let us call $d^{(u)} = (w^{(1,u)} ,w^{(2,u)}$) an upper double walk
and $d^{(r)} = (w^{(1,r)} ,w^{(2,r)}$) a right double walk. But for an unambiguous restoration of the minimumal double walk $d$ it is not enough to know these pieces.
It is also necessary to know the multiplicity of the edge $(r,v)$ and the behavior of the double walk $d$ at the vertices $r$ and $v$ (that is, after what moments of passing $r$ and $v$ the edge  $(r,v)$ passed). Let's call this information code of the double walk $d$. Thus, after removing the edge $(r,v)$ from $G$, instead of  one set of double walks we get a set of upper double walks, a set of right double walks and a set of codes. We divide the original set of double walks into such non-intersecting subsets for which the corresponding set of upper double walks, the set of right double walks and the set of codes are independent (that is, there is a bijection between the original set of double walks and the cartesian product of the set of upper double walks, the set of right double walks and the set of codes). Then, using the  weight's multiplicity, we can write  a total weight of double walks from original set as a product of a total weight of upper double walks
by a total weight of right double walks by some number specified below. Then we do the same for new sets of double walks until the system of recurrence
relations closes. Each such step is carried out in two stages: (i) cut the graph $G_d$  along the root $r$. Let us call it the first cutting. (ii) The resulting piece of $G_d$, which contains the edge $(r,v)$, is split
along the vertex $v$. Let us call it the second cutting. Let us  introduce some notations. The first walk of a minimal double walk is called gray one, and the second --- blue one. The first vertex of the gray (blue) walk is called a gray (blue) root. One can see that $r$ is a gray root. We also denote the left graph $G_l=(r,v)\cup G_u$. $G_l$ is a tree with the root $r$ and exactly one edge extending from the root --- $(r,v)$. Half of the length of the gray (blue) walk we denote $l_g$ ($l_b$). Let also $u_g$ ($u_b$) denotes half of the length of a gray (blue) walk along the upper graph, and $f_g$ ($f_b$) denotes the number of gray (blue) steps from the gray root $r$ to the vertex $v$. We also denote by $ r_g $ ($ r_b $) the number of all gray (blue) steps leaving  gray root $r$, and denote by  $v_g$ ($ v_b$) the number of steps leaving $v$ vertex, other than $\overrightarrow {(v,r)}$.

Let $\mathrm{Set}(l_g,l_b)$ denote the set of essential $(l_g,l_b)$-walks, and $\mathrm {S}(l_g,l_b)$ denote their total weight.  The following table describes the used denotations. The same denotations are also used for total weight $S$.

\begin{center}\label{tabular}
\begin{tabular}{|r|l|}
\hline
 $\mathrm{Set}^{(a)}$ & \textrm{ the
parameters } $r_g(d)$,
$r_b(d)$ \textrm{in this class can take any valid }\\
& \textrm{values}\\
\hline \textrm{the absence of} $^{(a)}$ &\textrm{the parameters } $r_g(d)$,
$r_b(d)$ \textrm{are fixed}\\
 \hline
  $\mathrm{Set}_{(=)}$ &
\textrm{
the gray root matches the blue one}  \\
\hline
$\mathrm{Set}_{(\neq)}$  & \textrm{the gray root doesn't match the blue one } \\ \hline $\mathrm{Set}_{(c)}$  & \textrm{ the skeleton of gray walk and skeleton of blue walk have at least one common edge }\\ \hline
$\mathrm{Set}_{(\not{c})}$  & \textrm{ the skeleton of gray walk and skeleton of blue walk have at least zero common edge}\\
 \hline
  $\mathrm{Set}^{(r)}$  & \textrm{the skeleton of blue walk have the edge}
$(r,v)$ \\
 \hline
$\mathrm{Set}^{(g)}$  & \textrm{the skeleton of blue walk don't have  the edge}
$(r,v)$\\
\hline
 $\mathrm{Set}^{(u)}$  & \textrm{ the blue root is in the upper tree}\\
\hline
 $\mathrm{Set}^{(d)}$
& \textrm{
the blue root which does not coincide with the gray root is in the right tree}\\
\hline
 $\mathrm{Set}^{(v)}$  &
\textrm{the parameters} $v_g(d)$, $v_b(d)$ \textrm{are fixed}\\
\hline
$\mathrm{Set}^{(1)}$  &
 \textrm{gray or blue walk is lacking}\\
 \hline
  $\mathrm{Set}_{(1)}$  & \textrm{the skeleton of the double walk $G_{d}$ has only one edge with the gray root} $r$\\
\hline
$\mathrm{Set}^{(l)}$  & \textrm{the lengths of the gray walk  and blue one on the upper (left) graph }\\
& \textrm{are fixed}\\
\hline
 $\mathrm{Set}^{(f)}$
& \textrm{ gray multiplicity and blue multiplicity of the edge }  $(r,v)$
\textrm{ are fixed}\\
 \hline
  $\mathrm{Set}^{(\varnothing)}$  &
\textrm{top graph is empty}\\
\hline
  $\mathrm{Set}^{(s)}$  &
\textrm{blue walk visits the gray root}\\
\hline
$\mathrm{Set}^{(n)}$  &
\textrm{gray walk doesn't have any steps}\\
\hline
 $\mathrm{{}^{(1)}Set}$  &
\textrm{gray root is in the first component ($V_{1,d}$) }\\
\hline
 $\mathrm{{}^{(2)}Set}$  &
\textrm{gray root is in the second component ($V_{2,d}$) }\\
\hline
 \end{tabular}
\end{center}

\newpage
\begin{figure}[ht]
\centering
\includegraphics[width=6in,height=5.175in]{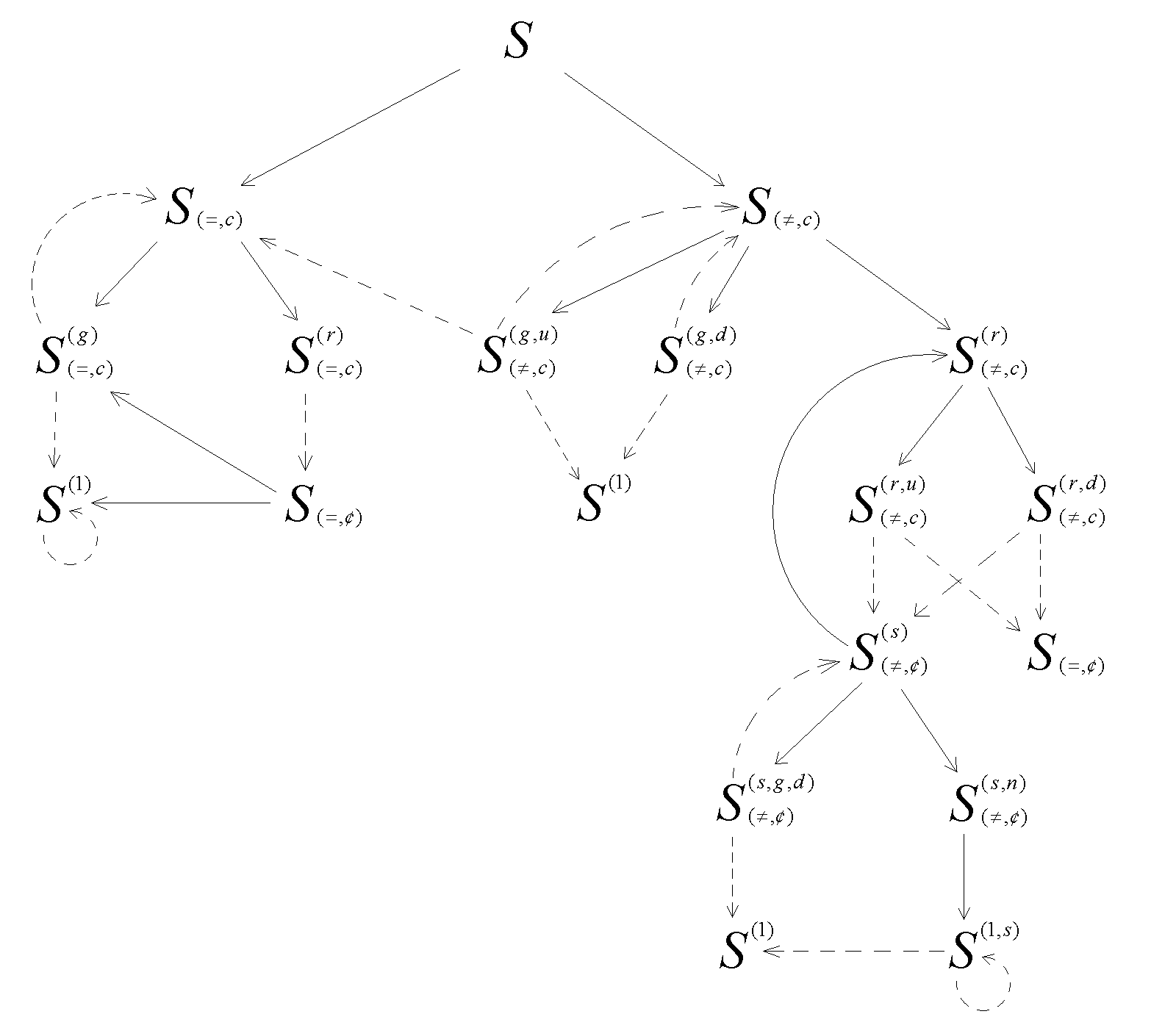}
\caption{Scheme of the system of recurrence relations}\label{scheme}
\end{figure}

Schematically, the system of recurrence relations is presented in Figure \ref{scheme}.
 Each element of the scheme is expressed through those elements which are indicated by arrows coming from it.
 The dotted arrow means that the total length is necessarily reduced.

 In the figures, the blue root is depicted as a black circle, the gray one is depicted as a white circle, and if the gray and blue roots coincide, the circle will be black and white. Two parallel segments indicate gray and blue edge. The case when the blue walk reaches the
   gray root is depicted as a small black circle inside the gray root.

Since for each essential double walk the gray and blue roots either coincide or do not coincide, the following equality is true.
\begin{equation}\label{rec}
 n^{(p,\alpha)}_{l_g,l_b}=\mathrm{S}
(l_g,l_b;p,\alpha)=\mathrm{S}_{( =,\, c)}^{(a)}(l_g,l_b;p,\alpha)+\mathrm{S}_{(\not
=,\,  c)}^{(a)}(l_g,l_b;p,\alpha).
\end{equation}
$\mathrm{S}
(l_g,l_b;p,\alpha)$, $\mathrm{S}_{( =,\, c)}^{(a)}(l_g,l_b;p,\alpha)$, $\mathrm{S}_{(\not
=,\,  c)}^{(a)}(l_g,l_b;p,\alpha)$ and other $\mathrm{S}$ depend on $p$ and $\alpha$, but in order not to overload the formulas, we will further omit the explicit expression of this dependence.
Looking through all possible values of the parameters $r_g$ and $r_b$ we get the following equality
\begin{equation}\label{rec_a=c}
   \mathrm{S}_{( =,\,
c)}^{(a)}(l_g,l_b)=\sum_{r_g=0}^{l_g} \sum_{r_b=0}^{l_b}
\mathrm{S}_{(=,\, c)}(l_g,l_b;r_g,r_b) .
\end{equation}
Since the edge $(r,v)$ is either in the blue skeleton or not, the following equality holds.
\begin{equation}\label{rec_=c}
  \mathrm{S}_{(=,\, c)}(l_g,l_b;r_g,r_b)= \mathrm{S}_{(=,\,
c)}^{(g)}(l_g,l_b;r_g,r_b) + \mathrm{S}_{(=,\,
c)}^{(r)}(l_g,l_b;r_g,r_b)
\end{equation}
Since the grey root  is either in the first component or in the second one, the following equality holds.
\begin{equation}\label{rec_=c}
  \mathrm{S}_{(=,\, c)}(l_g,l_b;r_g,r_b)= {}^{(1)}\mathrm{S}_{(=,\, c)}(l_g,l_b;r_g,r_b) + {}^{(2)}\mathrm{S}_{(=,\, c)}(l_g,l_b;r_g,r_b)
\end{equation}
Take an arbitrary double bipartite walk $d$ from $\mathrm{Set}_{(=,\,c)}^{(g)}(l_g,l_b;r_g,r_b)$. We divide its skeleton $G_{d}$ into the left graph $G_l$ and right one $G_r$. And the double bipartite walk $d$ breaks into a left double bipartite walk $f$ and a right one $s$. At the same time, $f$ is really a single walk, since there is no blue walk in $f$ (the edge $(r,v)$ in the blue walk $ d $ is not traversed, and the skeleton $G_{d}$ is a  tree). At the  root $r$ of the skeleton of $f$ there is only one edge, therefore $ f \in \mathrm{Set}^{(1)}_{(1)}(f_g+u_g,f_g)$. Once in $G_{d}$ there is a blue-gray edge, but in $G_l$ it does not exist, then it is in $G_r$. The gray and blue root in $s$ correspond, therefore $ s \in \mathrm{Set}_{(=,c)}(l_g-u_g-f_g,l_b,r_g-f_g,r_b)$.
The following lemma holds:

\begin{lemma}[The first cutting lemma]\label{l1}
Let $l_g, l_b, r_g, r_b$ be natural numbers  such that $l_g\geq r_g> 0$ and $l_b\geq r_b> 0$. Then the following equalities are true.
\begin{equation}\label{l1_1}
 {}^{(1)}\mathrm{S}_{(=,\,c)}^{(g)}(l_g,l_b;r_g,r_b)\!=\!\sum_{u_g=0}^{l_g-r_g}\sum_
 {f_g=1}^{r_g}{}^{(1)}\mathrm{S}^{(g,l,f)}_{(=,\,c)}(l_g,l_b;r_g,r_b;u_g,f_g)
\end{equation}
\begin{multline}\label{l1_2.0}
{}^{(1)}\mathrm{S}^{(g,l,f)}_{(=,c)}(l_g,l_b;r_g,r_b;u_g,f_g)\!=\!
\\ \!=\! \alpha_{1}^{-1}\cdot\binom{r_g-1}{f_g-1}\cdot
{}^{(1)}\mathrm{S}^{(1)}_{(1)}(f_g+u_g,f_g) \cdot {}^{(1)}\mathrm{S}_{(=,c)}(l_g-u_g-f_g,l_b,r_g-f_g,r_b)
\end{multline}
\begin{equation}\label{l1_1.1}
 {}^{(2)}\mathrm{S}_{(=,\,c)}^{(g)}(l_g,l_b;r_g,r_b)\!=\!\sum_{u_g=0}^{l_g-r_g}\sum_
 {f_g=1}^{r_g}{}^{(2)}\mathrm{S}^{(g,l,f)}_{(=,\,c)}(l_g,l_b;r_g,r_b;u_g,f_g)
\end{equation}
\begin{multline}\label{l1_2}
{}^{(2)}\mathrm{S}^{(g,l,f)}_{(=,c)}(l_g,l_b;r_g,r_b;u_g,f_g)\!=\!
\\ \!=\! \alpha_{2}^{-1}\cdot\binom{r_g-1}{f_g-1}\cdot
{}^{(2)}\mathrm{S}^{(1)}_{(1)}(f_g+u_g,f_g) \cdot {}^{(2)}\mathrm{S}_{(=,c)}(l_g-u_g-f_g,l_b,r_g-f_g,r_b)
\end{multline}
\end{lemma}
\begin{figure}[ht]
\centering
\includegraphics[width=3.963in,height=1.5in]{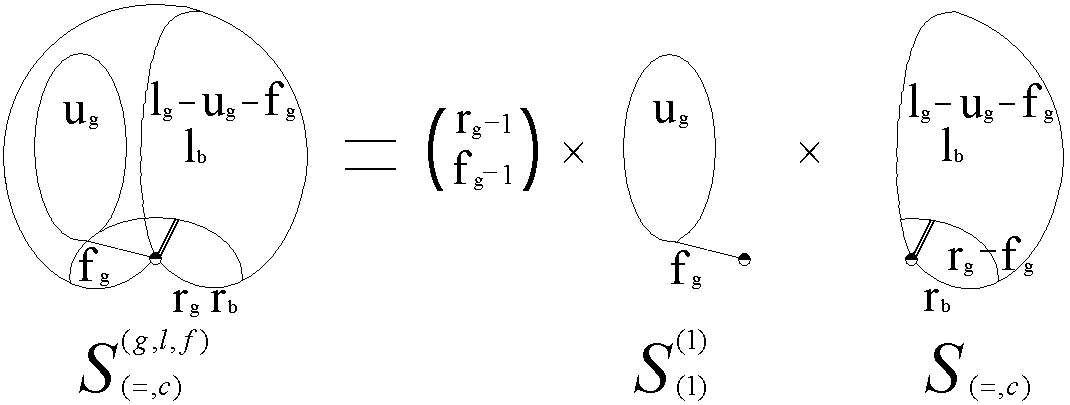}
\caption{ Representation of  ${}^{(1)}\mathrm{Set}^{(g,l,f)}_{(=,c)}$}\label{Image_1}
\end{figure}

What is ${}^{(1)}\mathrm{Set}_{(=,\,c)}^{(g)}(l_g,l_b;r_g,r_b)$? If we look at the spreadsheet we can easily understand it. ${}^{(1)}\mathrm{Set}_{(=,\,c)}^{(g)}(l_g,l_b;r_g,r_b)$ is a set
of essential double bipartite closed walks such that: 1) the length of the first(gray) walk is $2l_g$, the length of the second(blue) walk is $2l_b$, 2) ${}^{(1)}$ means that the root of the first walk is 1 and first step of the first walk is $\overrightarrow{(1,\tilde{1})}$, 3) ${}_{=}$ means that the root of the second walk coincides with the root of the first root, so it's 1, 4) ${}_{c}$ means that the skeleton of the first walk and the skeleton of the second walk have at least one common edge, 5)${}^{g}$ means that the skeleton of the second walk doesn't have edge $(1,\tilde{1})$, so $f_b=0$, then $u_b=0$, because the skeleton of the second walk is a tree and doesn't have edge $ (1,\tilde{1}$, but does have vertex 1, 6) the number steps of the first walk from vertex 1 is $r_g$, the number steps of the second walk from vertex 1 is $r_b$. So going over all the possible values $u_g, u_b$, we get \ref{l1_1}.
\ref{l1_2.0} follows from
 the following bijection:
\begin{multline}\label{bij1}
{}^{(1)}\mathrm{Set}^{(g,\,l,\,f)}_{(=,\,c)}(l_g,l_b;r_g,r_b;u_g,f_g)\!\to\!
{}^{(1)}\mathrm{Set}^{(1)}_{(1)}(f_g+u_g,f_g) \times \\
{}^{(1)}\mathrm{Set}_{(=,\,c)}(l_g-u_g-f_g,l_b,r_g-f_g,r_b)\times \mathrm{Code}^{(1)}(r_g,f_g),
\end{multline}
where
$ \mathrm{Code}^{(1)}(r_g,f_g)$
  is a set of sequences of zeros and ones of length $r_g$, which have exactly $f_g$ ones and the first term is 1. Figure \ref{Image_1} illustrates the equality  $(\ref{l1_2.0})$.

Indeed,
since the contribution of essential double bipartite walks is multiplicative along the edges and vertices (see \ref{eq:ess}), the contribution of a essential walk from ${}^{(1)}\mathrm{Set}^{(g,l,f)}_{(=,c)}(l_g,l_b;r_g,r_b;u_g,f_g)$ is equal to the product of the contributions of its parts from ${}^{(1)}\mathrm{Set}^{(g,l,f)}_{(=,c)}(l_g,l_b;r_g,r_b;u_g,f_g)$ and ${}^{(1)}\mathrm{Set}_{(=,c)}(l_g-u_g-f_g,l_b,r_g-f_g,r_b)$ and a factor $\alpha_{1}^{-1}$.
The multiplier $\alpha_{1}^{-1}$ arises due to the double use of the root in the first and second double bipartite walks of the partition.
Applying this fact and the Cartesian product of the image of the abovedescribed bijective map, we obtain the following equality
\begin{multline*}
{}^{(1)}\mathrm{S}^{(g,\,l,\,f)}_{(=,\,c)}(l_g,l_b;r_g,r_b;u_g,f_g)= \left|
\mathrm{Code}^{(1)}(r_g,f_g) \right| \cdot
{}^{(1)}\mathrm{S}^{(1)}_{(1)}(f_g+u_g,f_g) \cdot \\
\cdot {}^{(1)}\mathrm{S}_{(=,\,c)}(l_g-u_g-f_g,l_b,r_g-f_g,r_b) \cdot \alpha_{1}^{-1},
\end{multline*}
 $\left| \mathrm{Code}^{(1)}(r_g,f_g) \right|=\binom{r_g-1}
 {f_g-1} $ is a trivial combinatorial
fact. Indeed, if the first element is fixed , then it remains to choose $ f_g-1 $ positions for the remaining ones among the $ r_g-1 $ free
 places.

 It remains to prove bijectivity of \ref{bij1}. At first we will receive two numerical double walks and a code. And then we minimize both double walks.
  Splitting is performed using the following splitting algorithm. We first break into a gray walk.
 Go along $ d^{(1)} $ and if the next step belongs to $G_l$,
 then we add it to the first gray walk $f^{(1)}$, otherwise we add it to the second gray walk $s^{(1)}$.
At the same time, if the next step begins with the gray root, then, if it is
$ (\overline{r,v}) $, then we assign 1 to the code, otherwise  we assign 0 to the code.
Obviously, the first  element of code is 1, since the first edge of the
gray walk is by definition $(\overline{r,v})$, and the total number of ones in the code is $f_g$.
 I.e. $\# \{i: c_i = 1 \} = f_g$, $ \# \{i: c_i = 0 \} = r_g-f_g \, \, \wedge c_1 = 1$.
 It is easy to see that $f^{(1)}$ and $s^{(1)}$ are really
bipartite walks (in particular for every adjacent edges origin of next edge coincides with end of previous one), closed ones (in particular both walks  the
same root $r$). Obviously, every edge  from the left graph
$G_l$, and from the right graph $G_r$ is traversed in the corresponding
walk $f^{(1)}$ or $s^{(1)}$  the same number of times as in a gray walk $d^{(1)}$, i.e.
$$
\forall e \in G_l \quad n_{d^{(1)}}(e)=n_{f^{(1)}}(e) \,\, \textrm{and}
\,\, \forall e \in G_r \quad n_{d^{(1)}(e)}=n_{s^{(1)}}(e) .
$$
The blue walk does not need to be broken, since it will be completely in $s^{(2)}$. I.e.
 $$
\forall e \in G_l \quad n_{d^{(2)}}(e)=n_{f^{(2)}}(e)=0 \,\, \textrm{and}
\,\, \forall e \in G_r \quad n_{d^{(2)}}(e)=n_{s^{(2)}}(e).
$$
Thus, the weight of the original double walk is equal to the product of the weights of the first and
the second partitioned double walks up to factor $\alpha_1^{-1}$.
 Now make them minimal by applying to
them minimization mapping. At the same time, the weight of walks will not change.

Here's an example of double bipartite closed walk $d=(d^{(1)},d^{(2)})$. The bipartite closed walk
$d^{(1)}=(1,\tilde{1},2,\tilde{2},2,\tilde{3},2,\tilde{2},2,\tilde{1},3,\tilde{1},3,\tilde{1},1,\tilde{4},4,\tilde{4},5,\tilde{4},6,\tilde{4},$ $1,\tilde{5},1,\tilde{4},1,\tilde{1},1,\tilde{5},1,\tilde{1},
1,\tilde{4},1,\tilde{1},3,$ $\tilde{1},1)$. $d^{(1)}$ is a grey walk. $d^{(2)}=(1,\tilde{6},7,\tilde{7},7,\tilde{6},7,\tilde{7},7,\tilde{6},,1,\tilde{5},8,\tilde{5},1,\tilde{4},1,\tilde{4},9,\tilde{4},5,\tilde{4},$ $1,\tilde{5},1)$. $d^{(2)}$ is a blue walk.
$r=1, v=\tilde{1}$. $d\in {}^{(1)}\mathrm{Set}^{(g,\,l,\,f)}_{(=,\,c)}(19,12;9,5;7,4)$. $G= \left(\left\{1,\tilde{1},2,\tilde{2},\tilde{3},3,\tilde{4},4,5,6,\right. \right.$ $\left.\tilde{5},8,\tilde{6},7,\tilde{7},9\right\},$ $\left\{(1,\tilde{1}),(\tilde{1},2),(2,\tilde{2}),(2,\tilde{3}),(\tilde{1},3),(1,\tilde{4}), (\tilde{4},4),
 (\tilde{4},5),(\tilde{4},6),(1,\tilde{5}),\right.$ \newline $\left.\left.(\tilde{5},8),(1,\tilde{6}),(\tilde{6},7),(7,\tilde{7}),(\tilde{4},9)\right\}\right)$. $G_u=(\{\tilde{1},2,\tilde{2},\tilde{3},3\},$ $\{(\tilde{1},2),(2,\tilde{2}),(2,\tilde{3}),(\tilde{1},3)\}$. $G_l=(\{1,\tilde{1},2,\tilde{2},\tilde{3},3\},\{(1,\tilde{1}),(\tilde{1},2),(2,\tilde{2}),(2,\tilde{3}),(\tilde{1},3)\}$. $G_r=\left(\left\{1,\tilde{4},4,5,6,\tilde{5},8,\tilde{6},\right.\right.$ $7,\tilde{7},9\},\{(1,\tilde{4}),(\tilde{4},4),
 (\tilde{4},5),(\tilde{4},6),(1,\tilde{5}),(\tilde{5},8),(1,\tilde{6}),$ $(\tilde{6},7),(7,\tilde{7}),(\tilde{4},9)\})$. $l_g=19$, $l_b=12$, $f_g=4$, $f_b=0$,  $r_g=9$, $r_b=5$, $u_g=7$,
 $u_b=0$, $v_g=4$, $v_b=0$. $d\to (\eta,\theta,c)$. $\eta=(\eta^{(1)},\eta^{(2)})$. $\theta=(\theta^{(1)},\theta^{(2)})$. $c\in\{0,1\}^9$. $\eta^{(1)}=(1,\tilde{1},2,\tilde{2},2,\tilde{3},$ $2,\tilde{2},2,\tilde{1},3,\tilde{1},3,\tilde{1},1,$ $\tilde{1},1,\tilde{1},1,\tilde{1},3,\tilde{1},1)$. $\eta^{(2)}=(1)$.
 $\theta^{(1)}=(1,\tilde{4},4,\tilde{4},5,\tilde{4},6,\tilde{4},1,\tilde{5},1,\tilde{4},1,\tilde{5},1,\tilde{4},1)$. $\theta^{(2)}=(1,\tilde{6},7,\tilde{7},7,\tilde{6},7,\tilde{7},7,\tilde{6},1,\tilde{5},8,\tilde{5},1,\tilde{4},1,\tilde{4},9,\tilde{4},5,\tilde{4},1,\tilde{5},1)$. $c=(1,0,0,0,1,0,1,0,1)$.
  $\eta\in {}^{(1)}\mathrm{Set}^{(1)}_{(1)}(11,7)$. $\theta\in {}^{(1)}\mathrm{Set}_{(=,\,c)}(8,12,5,5)$.

Bijectivity is proved by the following collection algorithm.
We will gradually renumber the vertices of the first and second
walks. The roots of the first and second double walks will be put in compliance
number 1.
  Let's go around these double walks. We start the construction from the root.
  If the next step of double walk under construction
ends at the root, then if  the next element of the code is 1, then
go on the first subwalk $f$, otherwise go on the second subwalk $s$. If the final vertex of the current step along o the subwalk have not yet  its own number in the large walk, then we will put in correspondence with it the largest number of the already completed vertices of the large walk in the corresponding component plus 1 in the corresponding set.
 The result is bipartite double walk from the required class. It is easy to see that splitting mapping and
 collection mapping are injective. It means that they are bijective, since the area of definition and
the area of values are finite.
It remains to break the gray walk $f^{(1)}$.

\begin{lemma}[Second cutting lemma]\label{l2}
Let $f_g$ be a  natural number and $u_g$ be  a  natural number or zero. Then the following equalities are true.
\begin{equation}\label{l2_1}
{}^{(1)}\mathrm{S}^{(1)}_{(1)}(f_g+u_g,f_g)=\sum_{v_g=0}^{u_g}
{}^{(1)}\mathrm{S}^{(1,\,v)}_{(1)}(f_g+u_g,f_g,v_g)
\end{equation}
\begin{equation}\label{l2_2}
{}^{(1)}\mathrm{S}^{(1,\,v)}_{(1)}(f_g+u_g,f_g,v_g)\!=\!
\binom{f_g+v_g-1}{f_g-1}\cdot \frac{\alpha_1\cdot V_{2f_g}}{p^{f_g-1}}
\cdot {}^{(2)}\mathrm{S}^{(1)}(u_g,v_g)
\end{equation}
\begin{equation}\label{l2_1.0}
{}^{(2)}\mathrm{S}^{(1)}_{(1)}(f_g+u_g,f_g)=\sum_{v_g=0}^{u_g}
{}^{(2)}\mathrm{S}^{(1,\,v)}_{(1)}(f_g+u_g,f_g,v_g)
\end{equation}
\begin{equation}\label{l2_2.0}
{}^{(2)}\mathrm{S}^{(1,\,v)}_{(1)}(f_g+u_g,f_g,v_g)\!=\!
\binom{f_g+v_g-1}{f_g-1}\cdot \frac{\alpha_2\cdot V_{2f_g}}{p^{f_g-1}}
\cdot {}^{(1)}\mathrm{S}^{(1)}(u_g,v_g)
\end{equation}
\end{lemma}
\begin{figure}[ht]
\centering
\includegraphics[width=3.5955in,height=1.5in]{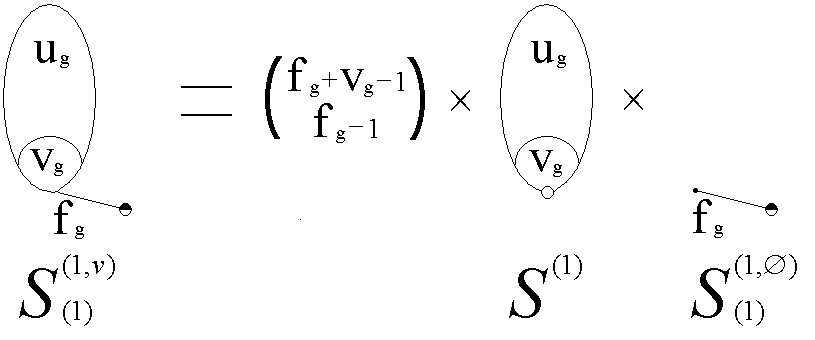}
\caption{Representation of ${}^{(1)}\mathrm{Set}^{(1,\,v)}_{(1)}$}\label{Image_2}
\end{figure}
This lemma is proved in the same way as the first one. The first equality is obvious, and the second follows from the following bijection
\begin{equation}\label{bij2}
{}^{(1)}\mathrm{Set}^{(1,v)}_{(1)}(f_g+u_g,f_g,v_g) \stackrel{bij}{\to}  {}^{(2)}\mathrm{Set}^{(1)}(u_g,v_g)
 \times {}^{(1)}\mathrm{Set}^{(1,\,\emptyset)}_1(f_g) \times \mathrm{ Code}^{(2)}(f_g+v_g,f_g),
\end{equation}
where
 $\mathrm{Code}^{(2)}(f_g+v_g,f_g)$
 is a set of sequences of zeros and ones of length $ f_g + v_g $, which have exactly $f_g$ ones and the last term is 1. The last term is 1, since the gray walk should return to the gray root $r$  by last step from the vertex $v$. ${}^{(1)}\mathrm{Set}^{(1,\emptyset)}_1 (f_g)$ consists of a single walk with a weight $\frac{\alpha_1\cdot\alpha_2\cdot V_{2f_g}}{p^{f_g-1}}$. Figure \ref{Image_2} illustrates the equality $(\ref{l2_2})$.

Combining these two lemmas, changing the order of summation and taking out some expressions beyond
  sign of the sum, we get the formulas
   $$
{}^{(1)}\mathrm{S}_{(=,\,  c)}^{(g)}(l_g,l_b;r_g,r_b)=\sum_{f_g=1}^{r_g} \binom{r_g-1}{f_g-1} \cdot
       \frac{ V_{2f_g}}{p^{f_g-1}}
       \cdot \sum_{u_g=0}^{l_g-r_g} {}^{(1)}\mathrm{S}_{(=,\,
       c)}(l_g-u_g-f_g,l_b;r_g-f_g,r_b)\cdot
 $$
\begin{equation}\label{rec_g=c}
      \cdot \sum_{v_g=0}^{u_g} \binom{f_g+v_g-1}{f_g-1} \cdot {}^{(2)}\mathrm{S}^{(1)}(u_g,v_g),
\end{equation}
   $$
{}^{(2)}\mathrm{S}_{(=,\,  c)}^{(g)}(l_g,l_b;r_g,r_b)=\sum_{f_g=1}^{r_g} \binom{r_g-1}{f_g-1} \cdot
       \frac{ V_{2f_g}}{p^{f_g-1}}
       \cdot \sum_{u_g=0}^{l_g-r_g} {}^{(2)}\mathrm{S}_{(=,\,
       c)}(l_g-u_g-f_g,l_b;r_g-f_g,r_b)\cdot
 $$
\begin{equation}\label{rec_g=c_1}
      \cdot \sum_{v_g=0}^{u_g} \binom{f_g+v_g-1}{f_g-1} \cdot {}^{(1)}\mathrm{S}^{(1)}(u_g,v_g).
\end{equation}
\subsection{Conclusion of a recursive system of equations}

\begin{figure}[ht]
\centering
\includegraphics[width=4.329in,height=1.5in]{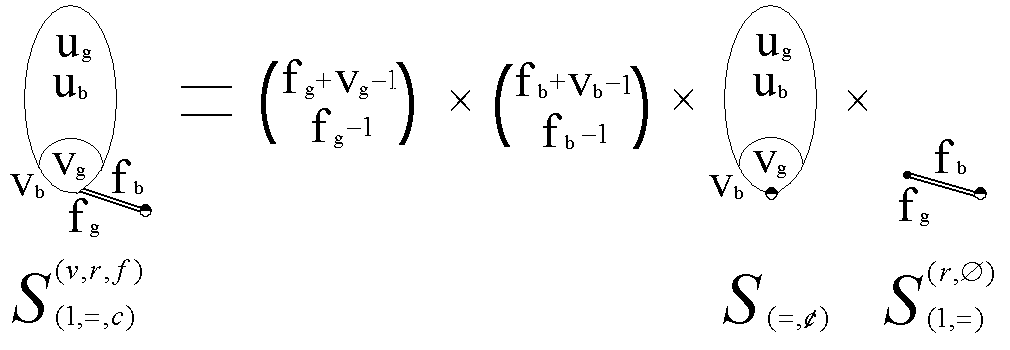}
\caption{Representation of ${}^{(1)}\mathrm{Set}^{(v,r,f)}_{(1,\,=,\,  c)}$}\label{Picture_2_2}
\end{figure}

Similarly (see also \cite{KSV}) the next formulas are proved.
$$
{}^{(1)}\mathrm{S}^{(1)}(l_g,r_g)=\sum_{f_g=1}^{r_g} \binom{r_g-1}{f_g-1}  \cdot \frac{ V_{2f_g}}{p^{f_g-1}}
        \cdot
$$
\begin{equation}\label{rec_1}
 \cdot\sum_{u_g=0}^{l_g-r_g} {}^{(1)}\mathrm{S}^{(1)}(l_g-u_g-f_g,r_g-f_g)
      \cdot \sum_{v_g=0}^{u_g} \binom{f_g+v_g-1}{f_g-1} \cdot
      {}^{(2)}\mathrm{S}^{(1)}(u_g,v_g),
\end{equation}
$$
{}^{(2)}\mathrm{S}^{(1)}(l_g,r_g)=\sum_{f_g=1}^{r_g} \binom{r_g-1}{f_g-1}  \cdot \frac{ V_{2f_g}}{p^{f_g-1}}
        \cdot
$$
\begin{equation}\label{rec_1_1}
 \cdot\sum_{u_g=0}^{l_g-r_g} {}^{(2)}\mathrm{S}^{(1)}(l_g-u_g-f_g,r_g-f_g)
      \cdot \sum_{v_g=0}^{u_g} \binom{f_g+v_g-1}{f_g-1} \cdot
      {}^{(1)}\mathrm{S}^{(1)}(u_g,v_g).
\end{equation}
 The formulas
$$
 {}^{(1)}\mathrm{S}_{(=,\,  c)}^{(r)}(l_g,l_b;r_g,r_b)=\sum_{f_g=1}^{r_g}  \binom{r_g-1}{f_g-1} \cdot
     \sum_{f_b=1}^{r_b} \binom{r_b}{f_b} \cdot \frac{
     V_{2(f_g+f_b)}}{p^{f_g+f_b-1}}
       \cdot
$$
$$
       \cdot \sum_{u_g=0}^{l_g-r_g} {}^{(1)}\mathrm{S}_{( =,\, \not
c)}(l_g-u_g-f_g,l_b-u_b-f_b;r_g-f_g,r_b-f_b) \, \cdot
$$
\begin{equation}\label{rec_r=c}
 \cdot \sum_{v_g=0}^{u_g} \binom{f_g+v_g-1}{f_g-1} \cdot \sum_{v_b=0}^{u_b}
      \binom{f_b+v_b-1}{f_b-1} \cdot {}^{(2)}\mathrm{S}_{( =,\, \not c)}(u_g,v_g),
\end{equation}
$$
 {}^{(2)}\mathrm{S}_{(=,\,  c)}^{(r)}(l_g,l_b;r_g,r_b)=\sum_{f_g=1}^{r_g}  \binom{r_g-1}{f_g-1} \cdot
     \sum_{f_b=1}^{r_b} \binom{r_b}{f_b} \cdot \frac{
     V_{2(f_g+f_b)}}{p^{f_g+f_b-1}}
       \cdot
$$
$$
       \cdot \sum_{u_g=0}^{l_g-r_g} {}^{(2)}\mathrm{S}_{( =,\, \not
c)}(l_g-u_g-f_g,l_b-u_b-f_b;r_g-f_g,r_b-f_b) \, \cdot
$$
\begin{equation}\label{rec_r=c_1}
 \cdot \sum_{v_g=0}^{u_g} \binom{f_g+v_g-1}{f_g-1} \cdot \sum_{v_b=0}^{u_b}
      \binom{f_b+v_b-1}{f_b-1} \cdot {}^{(1)}\mathrm{S}_{( =,\, \not c)}(u_g,v_g)
\end{equation}
 follow from the following lemmas \ref{l3.5} and \ref{l3.6}. 
 \begin{lemma}\label{l3.5}
 Let $l_g, l_b, r_g, r_b$ be natural numbers  such that $l_g\geq r_g> 0$ and $l_b\geq r_b> 0$. Then the following equalities are true.
 $$
 {}^{(1)}\mathrm{S}_{(=,\,c)}^{(r)}(l_g,l_b;r_g,r_b)\!=\!\sum_{u_g=0}^{l_g-r_g}\sum_
 {f_g=1}^{r_g}\sum_{u_b=0}^{l_b-r_b}\sum_
 {f_b=1}^{r_b}{}^{(1)}\mathrm{S}^{(r,l,f)}_{(=,\, c)}(l_g,l_b;r_g,r_b;u_g,u_b;f_g,f_b),
$$
$$
{}^{(1)}\mathrm{S}^{(r,l,f)}_{(=,\, c)}(l_g,l_b;r_g,r_b;u_g,u_b;f_g,f_b)\!=\!
\alpha_1^{-1}\binom{r_g-1}{f_g-1} \cdot\binom{r_b}{f_b}\cdot
{}^{(1)}\mathrm{S}^{(r,f)}_{(1,\, =,\, c)}(f_g+u_g,f_b+u_b;f_g,f_b) \cdot
$$
$$
\cdot{}^{(1)}\mathrm{S}_{(=,\, \not c)}
(l_g-u_g-f_g,l_b-u_b-f_b;r_g-f_g,r_b-f_b),
$$
 $$
 {}^{(2)}\mathrm{S}_{(=,\,c)}^{(r)}(l_g,l_b;r_g,r_b)\!=\!\sum_{u_g=0}^{l_g-r_g}\sum_
 {f_g=1}^{r_g}\sum_{u_b=0}^{l_b-r_b}\sum_
 {f_b=1}^{r_b}{}^{(2)}\mathrm{S}^{(r,l,f)}_{(=,\, c)}(l_g,l_b;r_g,r_b;u_g,u_b;f_g,f_b),
$$
$$
{}^{(2)}\mathrm{S}^{(r,l,f)}_{(=,\, c)}(l_g,l_b;r_g,r_b;u_g,u_b;f_g,f_b)\!=\!
\alpha_2^{-1}\binom{r_g-1}{f_g-1} \cdot\binom{r_b}{f_b}\cdot
{}^{(2)}\mathrm{S}^{(r,f)}_{(1,\, =,\, c)}(f_g+u_g,f_b+u_b;f_g,f_b) \cdot
$$
$$
\cdot{}^{(2)}\mathrm{S}_{(=,\, \not c)}
(l_g-u_g-f_g,l_b-u_b-f_b;r_g-f_g,r_b-f_b).
$$
\end{lemma}
\begin{figure}[ht]
\centering
\includegraphics[width=4.053in,height=1.5in]{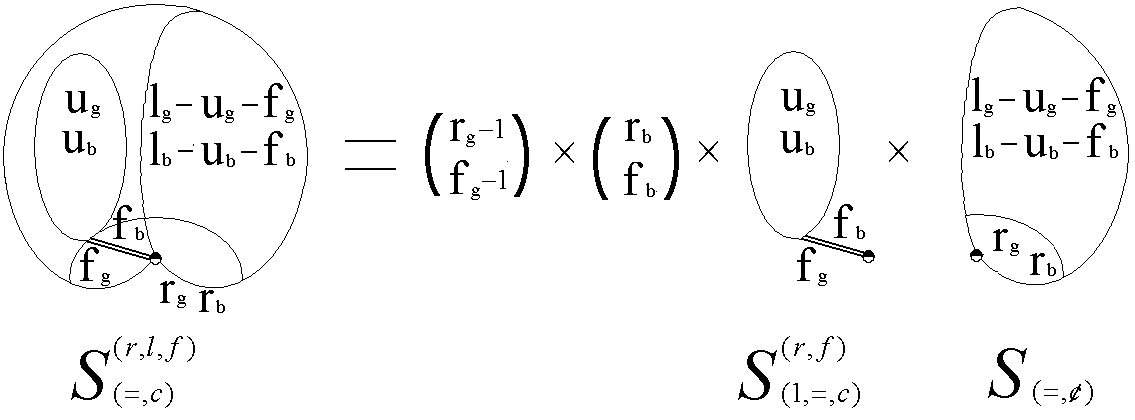}
\caption{Representation of ${}^{(1)}\mathrm{Set}_{(=,\,c)}^{(r,l,f)}$}\label{Picture_2_1}
\end{figure}
The formula contains the factor $\binom{r_b}{f_b}$, because, unlike the gray walk, the first step of the blue walk does not have to be $(r,v)$ (see Fig. \ref{Picture_2_1}).
\begin{lemma}\label{l3.6}
Let $f_g, f_b$ be   natural numbers and $u_g, u_b$ be    natural numbers or zero. Then the following equalities are true.
$$
{}^{(1)}\mathrm{S}^{(r,f)}_{(1,\, =,\, c)}(f_g+u_g,f_b+u_b;f_g,f_b) =\sum_{v_g=0}^{u_g}\sum_{v_b=0}^{u_b}
{}^{(1)}\mathrm{S}^{(v,r,f)}_{(1,\,=,\,  c)}(f_g+u_g,f_b+u_b;f_g,f_b;v_g,v_b),
$$
\begin{multline*}
{}^{(1)}\mathrm{S}^{(v,r,f)}_{(1,\,=,\,  c)}(f_g+u_g,f_b+u_b;f_g,f_b;v_g,v_b)\!=\! \\
\!=\!\binom{f_g+v_g-1}{f_g-1}\cdot\binom{f_b+v_b-1}{f_b-1}\cdot
 \frac{\alpha_1\cdot V_{2\cdot(f_g+f_b)}}{p^{f_g+f_b-1}}
\cdot {}^{(2)}\mathrm{S}_{(=,\, \not c)}(u_g,u_b;v_g,v_b).
\end{multline*}
The last formula is illustrated in Figure \ref{Picture_2_2}.
$$
{}^{(2)}\mathrm{S}^{(r,f)}_{(1,\, =,\, c)}(f_g+u_g,f_b+u_b;f_g,f_b) =\sum_{v_g=0}^{u_g}\sum_{v_b=0}^{u_b}
{}^{(2)}\mathrm{S}^{(v,r,f)}_{(1,\,=,\,  c)}(f_g+u_g,f_b+u_b;f_g,f_b;v_g,v_b),
$$
\begin{multline*}
{}^{(2)}\mathrm{S}^{(v,r,f)}_{(1,\,=,\,  c)}(f_g+u_g,f_b+u_b;f_g,f_b;v_g,v_b)\!=\! \\
\!=\!\binom{f_g+v_g-1}{f_g-1}\cdot\binom{f_b+v_b-1}{f_b-1}\cdot
 \frac{\alpha_2\cdot V_{2\cdot(f_g+f_b)}}{p^{f_g+f_b-1}}
\cdot {}^{(1)}\mathrm{S}_{(=,\, \not c)}(u_g,u_b;v_g,v_b).
\end{multline*}
\end{lemma}

\begin{figure}[ht]
\centering
\includegraphics[width=3.7326in,height=1.5in]{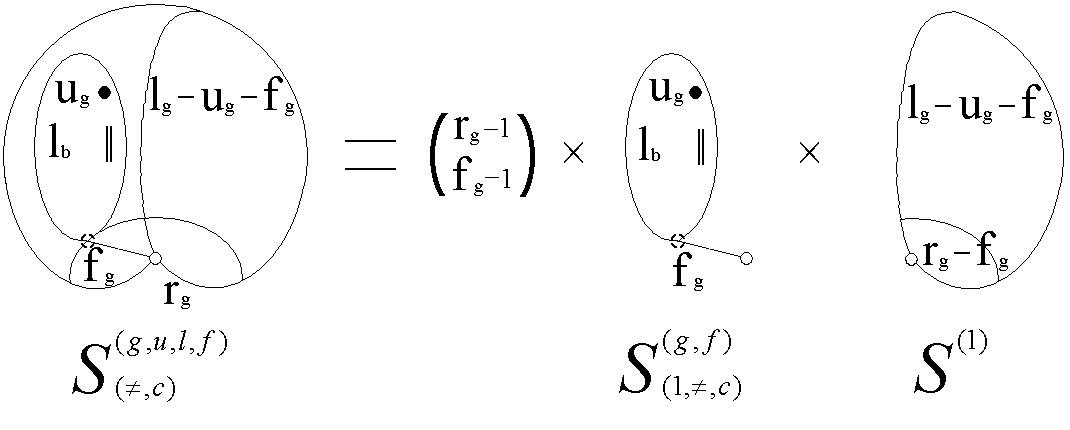}
\caption{Representation of ${}^{(1)}\mathrm{Set}^{(g,\,u,\,l,\,f)}_{(\not=,\,c)}$}\label{Picture_3_1}
\end{figure}

\begin{figure}[ht]
\centering
\includegraphics[width=3.7326in,height=1.5in]{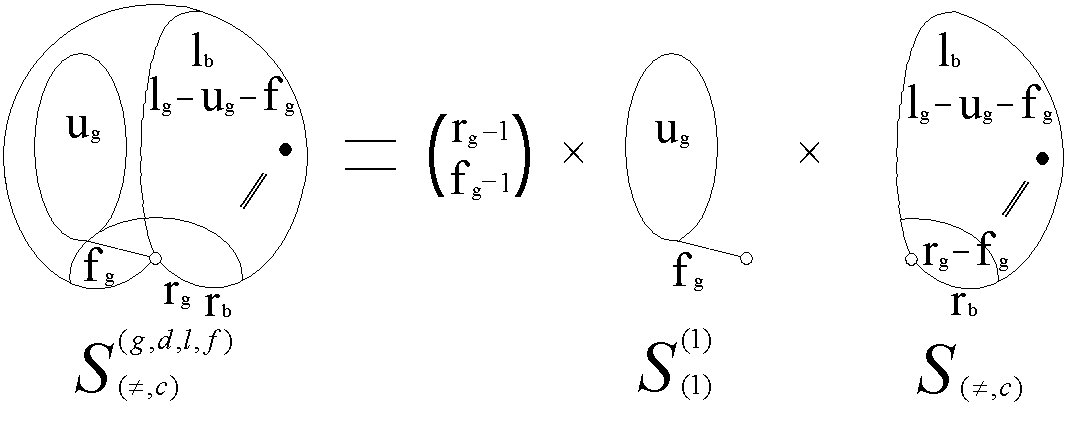}
\caption{Representation of ${}^{(1)}\mathrm{Set}^{(g,\,d,\,l,\,f)}_{(\not
=,\,c)}$}\label{Picture_4}
\end{figure}

\begin{figure}[ht]
\centering
\includegraphics[width=4.5732in,height=1.5in]{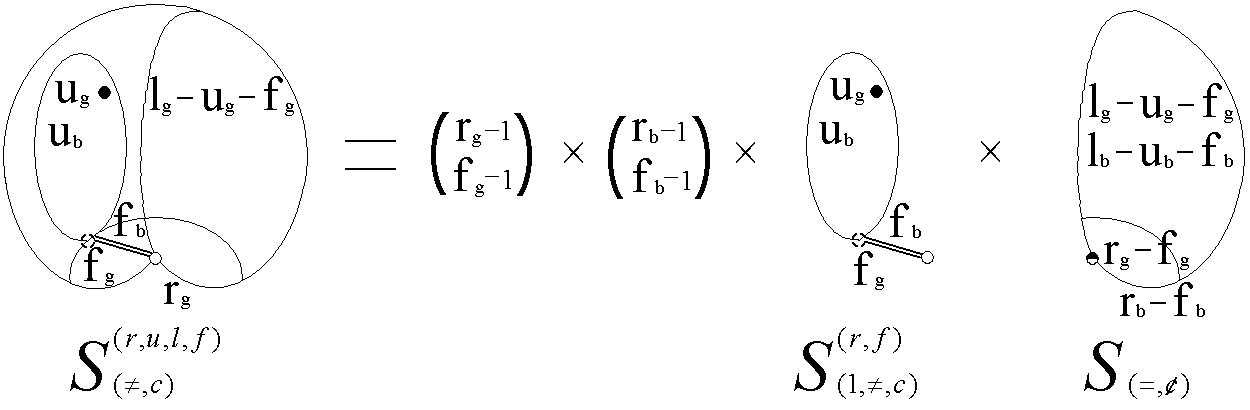}
\caption{Representation of ${}^{(1)}\mathrm{Set}^{(r,u,l,f)}_{(\not =,\, c)}$}\label{Picture_5_1}
\end{figure}

In a double bipartite walk from $ \mathrm{S}_{( =,\, \not c)}(l_g,l_b;r_g,r_b)$
either there is a blue-gray edge or there is not a blue-gray edge. In the first case, it is from
 $\mathrm{S}_{( =,\,  c)}(l_g,l_b;r_g,r_b)$. And in the second, the blue walk and gray walk  do not have common vertices except the gray root $r$, therefore they are practically independent. This implies the following equations
 \begin{equation}\label{rec_=nc}
 {}^{(1)}\mathrm{S}_{( =,\, \not c)}(l_g,l_b;r_g,r_b)= {}^{(1)}\mathrm{S}_{(=,\,  c)}(l_g,l_b;r_g,r_b)
  + \alpha_1^{-1}\cdot {}^{(1)}\mathrm{S}^{(1)}(l_g,r_g) \cdot {}^{(1)}\mathrm{S}^{(1)}(l_b,r_b),
\end{equation}
\begin{equation}\label{rec_=nc_1}
 {}^{(2)}\mathrm{S}_{( =,\, \not c)}(l_g,l_b;r_g,r_b)= {}^{(2)}\mathrm{S}_{(=,\,  c)}(l_g,l_b;r_g,r_b)
  +\alpha_2^{-1}\cdot {}^{(2)}\mathrm{S}^{(1)}(l_g,r_g) \cdot {}^{(1)}\mathrm{S}^{(2)}(l_b,r_b).
\end{equation}
Going over all possible values of the parameters $r_g$ and $r_b$, we deduce the equality
\begin{equation}\label{rec_an=c}
 \mathrm{S}_{(\not  =,\,  c)}^{(a)}(l_g,l_b)=\sum_{r_g=0}^{l_g} \sum_{r_b=0}^{l_b}
  \left({}^{(1)}\mathrm{S}_{(\not =,\, c)}(l_g,l_b;r_g,r_b)+{}^{(2)}\mathrm{S}_{(\not =,\, c)}(l_g,l_b;r_g,r_b)\right).
\end{equation}
In any essential double bipartite walk from,  either the edge $(r,v)$ is a blue-gray one or it's a pure gray one.
Therefore, the following equalities hold
\begin{equation}\label{rec_n=c}
{}^{(1)}\mathrm{S}_{(\not =,\, c)}(l_g,l_b;r_g,r_b)= {}^{(1)}\mathrm{S}_{(\not =,\, c)}^{(g)}
(l_g,l_b;r_g,r_b) + {}^{(1)}\mathrm{S}_{(\not =,\,  c)}^{(r)}(l_g,l_b;r_g,r_b).
\end{equation}
\begin{equation}\label{rec_n=c_1}
{}^{(2)}\mathrm{S}_{(\not =,\, c)}(l_g,l_b;r_g,r_b)= {}^{(2)}\mathrm{S}_{(\not =,\, c)}^{(g)}
(l_g,l_b;r_g,r_b) + {}^{(2)}\mathrm{S}_{(\not =,\,  c)}^{(r)}(l_g,l_b;r_g,r_b).
\end{equation}
If the blue root does not coincide with the gray one, then the blue root is either in the upper graph or lower one.
\begin{equation}\label{rec_gn=c}
{}^{(1)}\mathrm{S}_{(\not =,\, c)}^{(g)}(l_g,l_b;r_g,r_b)={}^{(1)}\mathrm{S}_{(\not =,\, c)}
^{(g,\, u)}(l_g,l_b;r_g,r_b) + {}^{(1)}\mathrm{S}_{(\not =,\, c)}^{(g,d)}(l_g,l_b;r_g,r_b)
\end{equation}
\begin{equation}\label{rec_gn=c_1}
{}^{(2)}\mathrm{S}_{(\not =,\, c)}^{(g)}(l_g,l_b;r_g,r_b)={}^{(2)}\mathrm{S}_{(\not =,\, c)}
^{(g,\, u)}(l_g,l_b;r_g,r_b) + {}^{(2)}\mathrm{S}_{(\not =,\, c)}^{(g,d)}(l_g,l_b;r_g,r_b)
\end{equation}
For a double walk from $ \mathrm{S}_{(\not =,\, c)}^{(g,\, u)}(l_g,l_b;r_g,r_b)$
since there is no edge $(r,v)$ in the blue skeleton, $r_b$ is 0. In the second double walk $s$ there is no blue component. We have the next lemma (see also Fig. \ref{Picture_3_1}).
 \begin{lemma} \label{l3.7}
 Let $l_g, l_b, r_g, r_b$ be natural numbers or zero such that $l_g\geq r_g> 0$ and $l_b\geq r_b \geq 0$. Then the following equalities are true.
$$
 {}^{(1)}\mathrm{S}_{(\not =,\, c)}^{(g,\, u)}(l_g,l_b;r_g,r_b)\!=\!\delta_{r_b}\cdot
 \sum_{u_g=0}^{l_g-r_g}\sum_
 {f_g=1}^{r_g}{}^{(1)}\mathrm{S}^{(g,u,l,f)}_{(\not =,\,c)}(l_g,l_b;r_g;u_g,f_g)
$$
\begin{multline*}
{}^{(1)}\mathrm{S}^{(g,\,u,\,l,\,f)}_{(\not =,\,c)}(l_g,l_b;r_g;u_g,f_g)
\!=\!\ \\ \!=\!\alpha_1^{-1}\cdot\binom{r_g-1}{f_g-1}\cdot
{}^{(1)}\mathrm{S}^{(g,\,f)}_{(1,\,\not =,\,c)}(f_g+u_g,l_b;f_g) \cdot
{}^{(1)}\mathrm{S}^{(1)}(l_g-u_g-f_g,r_g-f_g).
\end{multline*}
$$
 {}^{(2)}\mathrm{S}_{(\not =,\, c)}^{(g,\, u)}(l_g,l_b;r_g,r_b)\!=\!\delta_{r_b}\cdot
 \sum_{u_g=0}^{l_g-r_g}\sum_
 {f_g=1}^{r_g}{}^{(2)}\mathrm{S}^{(g,u,l,f)}_{(\not =,\,c)}(l_g,l_b;r_g;u_g,f_g)
$$
\begin{multline*}
{}^{(2)}\mathrm{S}^{(g,\,u,\,l,\,f)}_{(\not =,\,c)}(l_g,l_b;r_g;u_g,f_g)
\!=\!\ \\ \!=\!\alpha_2^{-1}\cdot\binom{r_g-1}{f_g-1}\cdot
{}^{(2)}\mathrm{S}^{(g,\,f)}_{(1,\,\not =,\,c)}(f_g+u_g,l_b;f_g) \cdot
{}^{(2)}\mathrm{S}^{(1)}(l_g-u_g-f_g,r_g-f_g).
\end{multline*}
\end{lemma}
If the blue root lies in the upper graph, then it either coincides with the vertex $v$ or it doesn't coincide with the vertex $v$. In the first case, a double walk along the upper graph  belongs to $\mathrm{S}_{(=,\,  c)}(u_g,l_b;v_g,v_b)$, and in the second case it belongs to $\mathrm{S}_{(\not =,\, c)}(u_g,l_b;v_g,v_b)$.
And  we have the next lemma.
\begin{lemma}\label{l3.8}
Let $f_g$ be a  natural number, $u_g$ be  a  natural number or zero, $l_b$ be a natural number or zero. Then the following equalities are true.
$$
{}^{(1)}\mathrm{S}^{(g,\,f)}_{(1,\,\not =,\,c)}(f_g+u_g,l_b;f_g)=\sum_{v_g=0}^{u_g}
\sum_{v_b=0}^{l_b}
{}^{(1)}\mathrm{S}^{(v,\,g,\,f)}_{(1,\,\not =,\,c)}(f_g+u_g,l_b;f_g;v_g)
$$
\begin{multline*}
{}^{(1)}\mathrm{S}^{(v,\,g,\,f)}_{(1,\,\not =,\,c)}(f_g+u_g,l_b;f_g;v_g)\!=\! \\
\!=\!\alpha_1\cdot\binom{f_g+v_g-1}{f_g-1}\cdot \frac{V_{2f_g}}{p^{f_g-1}}
\cdot     \left( {}^{(2)}\mathrm{S}_{(=,\,  c)}(u_g,l_b;v_g,v_b)+
{}^{(2)}\mathrm{S}_{(\not =,\, c)}(u_g,l_b;v_g,v_b) \right),
\end{multline*}
$$
{}^{(2)}\mathrm{S}^{(g,\,f)}_{(1,\,\not =,\,c)}(f_g+u_g,l_b;f_g)=\sum_{v_g=0}^{u_g}
\sum_{v_b=0}^{l_b}
{}^{(2)}\mathrm{S}^{(v,\,g,\,f)}_{(1,\,\not =,\,c)}(f_g+u_g,l_b;f_g;v_g)
$$
\begin{multline*}
{}^{(2)}\mathrm{S}^{(v,\,g,\,f)}_{(1,\,\not =,\,c)}(f_g+u_g,l_b;f_g;v_g)\!=\! \\
\!=\!\alpha_2\cdot\binom{f_g+v_g-1}{f_g-1}\cdot \frac{V_{2f_g}}{p^{f_g-1}}
\cdot     \left( {}^{(1)}\mathrm{S}_{(=,\,  c)}(u_g,l_b;v_g,v_b)+
{}^{(1)}\mathrm{S}_{(\not =,\, c)}(u_g,l_b;v_g,v_b) \right).
\end{multline*}
\end{lemma}
\begin{figure}[ht]
\centering
\includegraphics[width=5.0735in,height=1.5in]{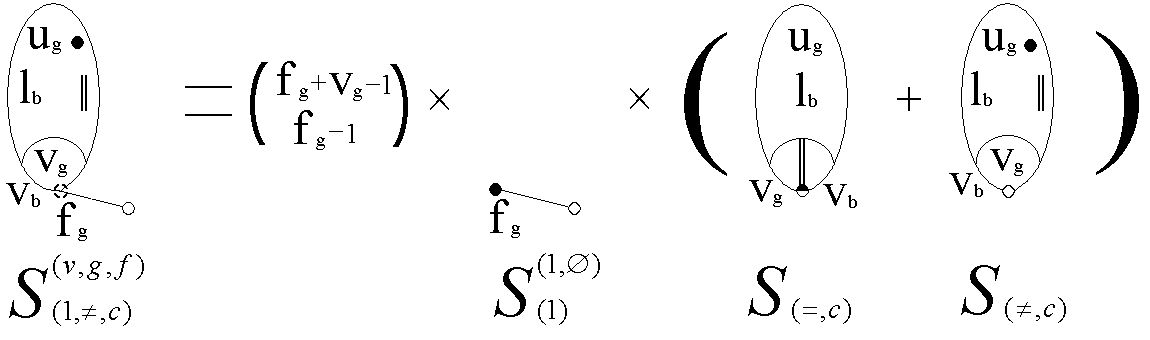}
\caption{Representation of ${}^{(1)}\mathrm{Set}^{(v,\,g,\,f)}_{(1,\,\not
=,\,c)}$}\label{Picture_3_2}
\end{figure}
The next equalities follow from   lemmas \ref{l3.7} and \ref{l3.8}.
$$
 {}^{(1)}\mathrm{S}_{(\not =,\, c)}^{(g,\, u)}(l_g,l_b;r_g,r_b)=\delta_{r_b} \cdot \sum_{f_g=1}^{r_g} \binom{r_g-1}{f_g-1}
      \cdot  \frac{ V_{2f_g}}{p^{f_g-1}}
       \cdot \sum_{u_g=0}^{l_g-r_g} {}^{(1)}\mathrm{S}^{(1)}(l_g-u_g-f_g,r_g-f_g)
      \cdot
$$
\begin{equation}\label{rec_gun=c}
 \cdot \sum_{v_g=0}^{u_g} \binom{f_g+v_g-1}{f_g-1} \cdot \sum_{v_b=0}^{l_b}
       \left( {}^{(2)}\mathrm{S}_{(=,\,  c)}(u_g,l_b;v_g,v_b)+{}^{(2)}\mathrm{S}_{(\not =,\, c)}
       (u_g,l_b;v_g,v_b) \right),
\end{equation}
$$
 {}^{(2)}\mathrm{S}_{(\not =,\, c)}^{(g,\, u)}(l_g,l_b;r_g,r_b)=\delta_{r_b} \cdot \sum_{f_g=1}^{r_g} \binom{r_g-1}{f_g-1}
      \cdot  \frac{ V_{2f_g}}{p^{f_g-1}}
       \cdot \sum_{u_g=0}^{l_g-r_g} {}^{(2)}\mathrm{S}^{(1)}(l_g-u_g-f_g,r_g-f_g)
      \cdot
$$
\begin{equation}\label{rec_gun=c_1}
 \cdot \sum_{v_g=0}^{u_g} \binom{f_g+v_g-1}{f_g-1} \cdot \sum_{v_b=0}^{l_b}
       \left( {}^{(1)}\mathrm{S}_{(=,\,  c)}(u_g,l_b;v_g,v_b)+{}^{(1)}\mathrm{S}_{(\not =,\, c)}
       (u_g,l_b;v_g,v_b) \right),
\end{equation}
The following formulas
 $$
 {}^{(1)}\mathrm{S}_{(\not =,\, c)}^{(g,d)}(l_g,l_b;r_g,r_b)=\sum_{f_g=1}^{r_g} \binom{r_g-1}{f_g-1}\cdot
         \frac{ V_{2f_g}}{p^{f_g-1}}
        \cdot \sum_{u_g=0}^{l_g-r_g} {}^{(1)}\mathrm{S}_{(\not =,\, c)}(l_g-u_g-f_g,l_b;r_g-f_g,r_b) \cdot
$$
\begin{equation}\label{rec_gdn=c}
 \cdot \sum_{v_g=0}^{u_g} \binom{f_g+v_g-1}{f_g-1} \cdot
 {}^{(2)}\mathrm{S}^{(1)}(u_g,v_g),
\end{equation}
 $$
 {}^{(2)}\mathrm{S}_{(\not =,\, c)}^{(g,d)}(l_g,l_b;r_g,r_b)=\sum_{f_g=1}^{r_g} \binom{r_g-1}{f_g-1}\cdot
         \frac{ V_{2f_g}}{p^{f_g-1}}
        \cdot \sum_{u_g=0}^{l_g-r_g} {}^{(2)}\mathrm{S}_{(\not =,\, c)}(l_g-u_g-f_g,l_b;r_g-f_g,r_b) \cdot
$$
\begin{equation}\label{rec_gdn=c_1}
 \cdot \sum_{v_g=0}^{u_g} \binom{f_g+v_g-1}{f_g-1} \cdot
 {}^{(1)}\mathrm{S}^{(1)}(u_g,v_g)
\end{equation}
 are obtained  from lemma \ref{l2} and lemma \ref{l3.9} (see also Fig. \ref{Picture_4}).
\begin{lemma}\label{l3.9}  Let $l_g$, $l_b$, $r_g$, $r_b$ be natural numbers or zero such that $l_g \geq r_g > 0$ and $l_b \geq r_b \geq 0$. Then the following equalities are true.
$$
{}^{(1)}\mathrm{S}_{(\not =,\, c)}^{(g,d)}(l_g,l_b;r_g,r_b)\!=\!\sum_{u_g=0}^{l_g-r_g}\sum_
 {f_g=1}^{r_g}{}^{(1)}\mathrm{S}^{(g,\,d,\,l,\,f)}_{(\not =,\,c)}(l_g,l_b;r_g,r_b;u_g,f_g)
$$
\begin{multline*}
{}^{(1)}\mathrm{S}^{(g,\,d,\,l,\,f)}_{(\not =,\,c)}(l_g,l_b;r_g,r_b;u_g,f_g)\!=\!
\\
=\alpha_1^{-1}\binom{r_g-1}{f_g-1}\cdot
{}^{(1)}\mathrm{S}^{(1)}_{(1)}(f_g+u_g,f_g) \cdot {}^{(1)}\mathrm{S}_{(\not =,\,c)}(l_g-u_g-f_g,l_b,r_g-f_g,r_b)
\end{multline*}
$$
{}^{(2)}\mathrm{S}_{(\not =,\, c)}^{(g,d)}(l_g,l_b;r_g,r_b)\!=\!\sum_{u_g=0}^{l_g-r_g}\sum_
 {f_g=1}^{r_g}{}^{(2)}\mathrm{S}^{(g,\,d,\,l,\,f)}_{(\not =,\,c)}(l_g,l_b;r_g,r_b;u_g,f_g)
$$
\begin{multline*}
{}^{(2)}\mathrm{S}^{(g,\,d,\,l,\,f)}_{(\not =,\,c)}(l_g,l_b;r_g,r_b;u_g,f_g)\!=\!
\\
=\alpha_2^{-1}\binom{r_g-1}{f_g-1}\cdot
{}^{(2)}\mathrm{S}^{(1)}_{(1)}(f_g+u_g,f_g) \cdot {}^{(2)}\mathrm{S}_{(\not =,\,c)}(l_g-u_g-f_g,l_b,r_g-f_g,r_b)
\end{multline*}
\end{lemma}
The blue root which does not coincide with the gray one can be located either in the upper graph  or lower (right) graph.
\begin{equation}\label{rec_rn=c}
{}^{(1)}\mathrm{S}_{(\not =,\, c)}^{(r)}(l_g,l_b;r_g,r_b)= {}^{(1)}\mathrm{S}_{(\not =,\, c)}^{(r,\, u)}
(l_g,l_b;r_g,r_b) + {}^{(1)}\mathrm{S}_{(\not =,\, c)}^{(r,\, d)}(l_g,l_b;r_g,r_b),
\end{equation}
\begin{equation}\label{rec_rn=c}
{}^{(2)}\mathrm{S}_{(\not =,\, c)}^{(r)}(l_g,l_b;r_g,r_b)= {}^{(2)}\mathrm{S}_{(\not =,\, c)}^{(r,\, u)}
(l_g,l_b;r_g,r_b) + {}^{(2)}\mathrm{S}_{(\not =,\, c)}^{(r,\, d)}(l_g,l_b;r_g,r_b).
\end{equation}
 One more lemma for  $\mathrm{S}_{(\not =,\, c)}^{(r,\, u)}(l_g,l_b;r_g,r_b)$ looks like this (see also Fig. \ref{Picture_5_1}):
 \begin{lemma}\label{l3.10} Let $l_g$, $l_b$, $r_g$, $r_b$ be natural numbers or zero such that $l_g \geq r_g > 0$ and $l_b \geq r_b > 0$. Then the following equalities are true.
$$
 {}^{(1)}\mathrm{S}_{(\not =,\, c)}^{(r,\, u)}(l_g,l_b;r_g,r_b)\!=\!\sum_{u_g=0}^{l_g-r_g}\sum_
 {f_g=1}^{r_g}\sum_{u_b=0}^{l_b-r_b}\sum_
 {f_b=1}^{r_b}{}^{(1)}\mathrm{S}^{(r,u,l,f)}_{(\not =,\, c)}(l_g,l_b;r_g,r_b;u_g,u_b;f_g,f_b)
$$
$$
{}^{(1)}\mathrm{S}^{(r,u,l,f)}_{(\not =,\, c)}(l_g,l_b;r_g,r_b;u_g,u_b;f_g,f_b)\!=\!
\binom{r_g-1}{f_g-1} \cdot\binom{r_b-1}{f_b-1}\cdot
{}^{(1)}\mathrm{S}^{(r,\,f)}_{(1,\,\not =,\, c)}(f_g+u_g,f_b+u_b;f_g,f_b) \cdot
$$
$$
\cdot\alpha_1^{-1}\cdot {}^{(1)}\mathrm{S}_{(=,\, \not c)}
(l_g-u_g-f_g,l_b-u_b-f_b;r_g-f_g,r_b-f_b).
$$
$$
 {}^{(2)}\mathrm{S}_{(\not =,\, c)}^{(r,\, u)}(l_g,l_b;r_g,r_b)\!=\!\sum_{u_g=0}^{l_g-r_g}\sum_
 {f_g=1}^{r_g}\sum_{u_b=0}^{l_b-r_b}\sum_
 {f_b=1}^{r_b}{}^{(2)}\mathrm{S}^{(r,u,l,f)}_{(\not =,\, c)}(l_g,l_b;r_g,r_b;u_g,u_b;f_g,f_b)
$$
$$
{}^{(2)}\mathrm{S}^{(r,u,l,f)}_{(\not =,\, c)}(l_g,l_b;r_g,r_b;u_g,u_b;f_g,f_b)\!=\!
\binom{r_g-1}{f_g-1} \cdot\binom{r_b-1}{f_b-1}\cdot
{}^{(2)}\mathrm{S}^{(r,\,f)}_{(1,\,\not =,\, c)}(f_g+u_g,f_b+u_b;f_g,f_b) \cdot
$$
$$
\cdot\alpha_2^{-1}\cdot {}^{(2)}\mathrm{S}_{(=,\, \not c)}
(l_g-u_g-f_g,l_b-u_b-f_b;r_g-f_g,r_b-f_b).
$$
\end{lemma}
The last formula contains the factor $\binom{r_b-1}{f_b-1}$, since the last step of the blue walk from the gray root of $r$ should be $\overrightarrow{(r,v)}$.
 In the second double walk may or may not be blue-gray edges due to the fact that the edge $(r,v)$ is gray-blue.

\begin{figure}[ht]
\centering
\includegraphics[width=6.4412in,height=1.5in]{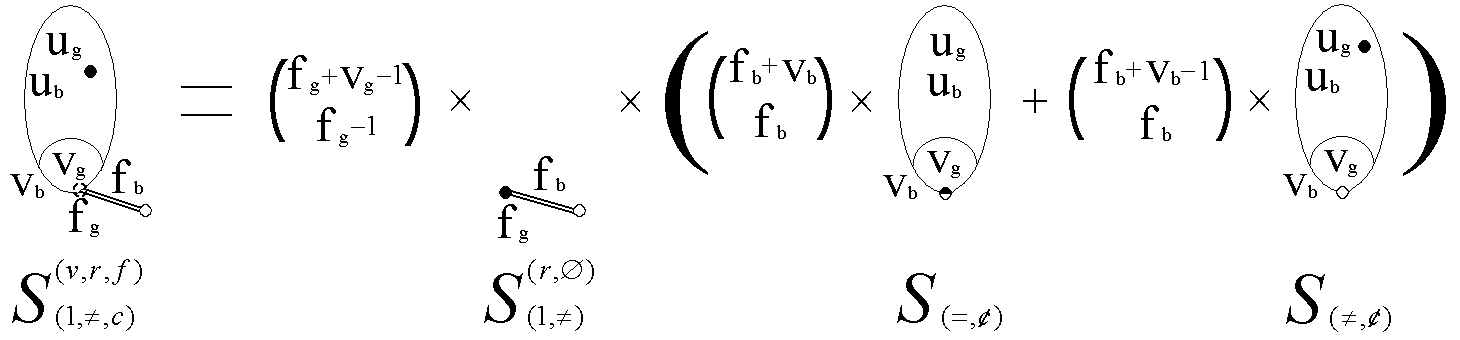}
\caption{Representation  ${}^{(1)}\mathrm{Set}^{(v,\,r,\,f)}_{(1,\,\not =,\,
c)}$}\label{Picture_5_2}
\end{figure}

  The following lemma for $\mathrm{S}_{(\not =,\, c)}^{(r,\, u)}(l_g,l_b;r_g,r_b)$ looks like this (see also Fig. \ref{Picture_5_2}) ~:
  \begin{lemma}\label{l3.11} Let $f_g, f_b$ be   natural numbers and $u_g, u_b$ be    natural numbers or zero. Then the following equalities are true.
  $$
{}^{(1)}\mathrm{S}^{(r,\,f)}_{(1,\,\not =,\, c)}(f_g+u_g,f_b+u_b;f_g,f_b) =\sum_{v_g=0}^{u_g}\sum_{v_b=0}^{u_b}
{}^{(1)}\mathrm{S}^{(v,\,r,\,f)}_{(1,\,\not =,\, c)}(f_g+u_g,f_b+u_b;f_g,f_b;v_g,v_b)
$$
$$
{}^{(1)}\mathrm{S}^{(v,\,r,\,f)}_{(1,\,\not =,\, c)}(f_g+u_g,f_b+u_b;f_g,f_b;v_g,v_b)\!=\!
\binom{f_g+v_g-1}{f_g-1}\cdot
 \frac{V_{2\cdot(f_g+f_b)}}{p^{f_g+f_b-1}}\,
\cdot\alpha_1\cdot
 $$
 $$
  \cdot \left( \binom{f_b+v_b}{f_b} \cdot{}^{(2)}\mathrm{S}_{( =,\,
\not c)}(u_g,u_b;v_g,v_b)+\binom{f_b+v_b-1}{f_b} \cdot
{}^{(2)}\mathrm{S}_{(\not =,\, \not c)}^{(s)}(u_g,u_b;v_g,v_b) \right)
$$
  $$
{}^{(2)}\mathrm{S}^{(r,\,f)}_{(1,\,\not =,\, c)}(f_g+u_g,f_b+u_b;f_g,f_b) =\sum_{v_g=0}^{u_g}\sum_{v_b=0}^{u_b}
{}^{(2)}\mathrm{S}^{(v,\,r,\,f)}_{(1,\,\not =,\, c)}(f_g+u_g,f_b+u_b;f_g,f_b;v_g,v_b)
$$
$$
{}^{(2)}\mathrm{S}^{(v,\,r,\,f)}_{(1,\,\not =,\, c)}(f_g+u_g,f_b+u_b;f_g,f_b;v_g,v_b)\!=\!
\binom{f_g+v_g-1}{f_g-1}\cdot
 \frac{V_{2\cdot(f_g+f_b)}}{p^{f_g+f_b-1}}\,
\cdot\alpha_2\cdot
 $$
 $$
  \cdot \left( \binom{f_b+v_b}{f_b} \cdot{}^{(1)}\mathrm{S}_{( =,\,
\not c)}(u_g,u_b;v_g,v_b)+\binom{f_b+v_b-1}{f_b} \cdot
{}^{(1)}\mathrm{S}_{(\not =,\, \not c)}^{(s)}(u_g,u_b;v_g,v_b) \right)
$$
\end{lemma}
The blue root either matches the vertex $v$ or not.
In the first case, a double walk along the upper graph will be from ${}^{(1)}\mathrm{S}_{(=,\, \not c)}(u_g,u_b;v_g,v_b)$, and in the second case,
 from ${}^{(1)}\mathrm{S}_{(\not=,\, \not c)}^{(s)}(u_g,u_b;v_g,v_b)$ (the blue walk along the upper graph should go along the vertex $v$, since the blue root
is in the upper graph, but the blue walk passes through the edge $(r,v)$).
Different factors are in the expression in parentheses, since in the second case the last step
from the vertex  $v$ have to not coincide with $\overrightarrow{(v,r)}$, but in
the first case it is optional. So we have the following formulas
$$
 {}^{(1)}\mathrm{S}_{(\not =,\, c)}^{(r,\, u)}(l_g,l_b;r_g,r_b)=\sum_{f_g=1}^{r_g} \binom{r_g-1}{f_g-1} \cdot
     \sum_{f_b=1}^{r_b} \binom{r_b-1}{f_b-1} \cdot \frac{ V_{2(f_g+f_b)}}{p^{f_g+f_b-1}}
                                 \cdot
$$
$$
  \cdot \sum_{u_g=0}^{l_g-r_g} \sum_{u_b=0}^{l_b-r_b}
         {}^{(1)}\mathrm{S}_{( =,\, \not c)}(l_g-u_g-f_g,l_b-u_b-f_b;r_g-f_g,r_b-f_b)
        \cdot \sum_{v_g=0}^{u_g} \binom{f_g+v_g-1}{f_g-1} \,\cdot
$$
\begin{equation}\label{rec_run=c}
 \cdot \sum_{v_b=0}^{u_b} \Bigl( \binom{f_b+v_b}{f_b} \cdot
        {}^{(2)}\mathrm{S}_{( =,\, \not c)}(u_g,u_b;v_g,v_b)+\binom{f_b+v_b-1}{f_b}
         \cdot {}^{(2}\mathrm{S}_{(\not =,\, \not c)}^{(s)}(u_g,u_b;v_g,v_b)
         \Bigl).
\end{equation}
$$
 {}^{(2)}\mathrm{S}_{(\not =,\, c)}^{(r,\, u)}(l_g,l_b;r_g,r_b)=\sum_{f_g=1}^{r_g} \binom{r_g-1}{f_g-1} \cdot
     \sum_{f_b=1}^{r_b} \binom{r_b-1}{f_b-1} \cdot \frac{ V_{2(f_g+f_b)}}{p^{f_g+f_b-1}}
                                 \cdot
$$
$$
  \cdot \sum_{u_g=0}^{l_g-r_g} \sum_{u_b=0}^{l_b-r_b}
         {}^{(2)}\mathrm{S}_{( =,\, \not c)}(l_g-u_g-f_g,l_b-u_b-f_b;r_g-f_g,r_b-f_b)
        \cdot \sum_{v_g=0}^{u_g} \binom{f_g+v_g-1}{f_g-1} \,\cdot
$$
\begin{equation}\label{rec_run=c_1}
 \cdot \sum_{v_b=0}^{u_b} \Bigl( \binom{f_b+v_b}{f_b} \cdot
        {}^{(1)}\mathrm{S}_{( =,\, \not c)}(u_g,u_b;v_g,v_b)+\binom{f_b+v_b-1}{f_b}
         \cdot {}^{(1)}\mathrm{S}_{(\not =,\, \not c)}^{(s)}(u_g,u_b;v_g,v_b)
         \Bigl).
\end{equation}

\begin{figure}[ht]
\centering
\includegraphics[width=3.9146in,height=1.5in]{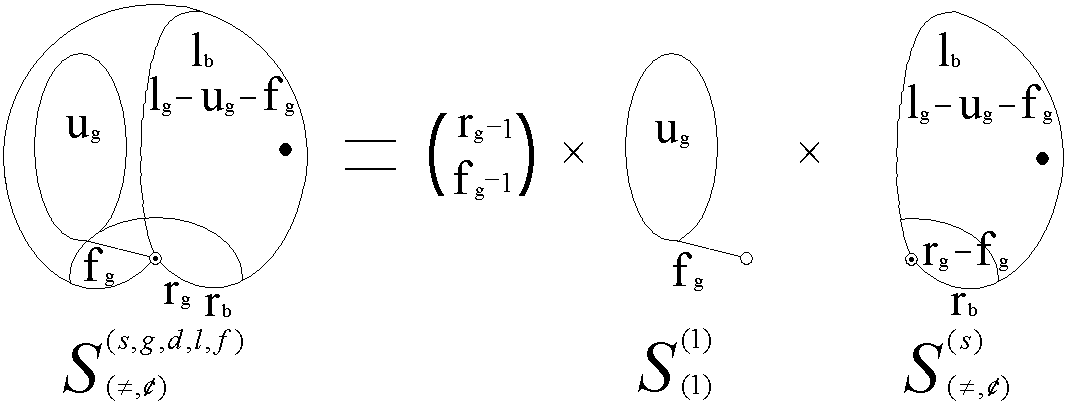}
\caption{Representation of ${}^{(1)}\mathrm{Set}^{(s,\,g,\,d\,,l,\,f)}_{(\not =,\,
\not c)}$}\label{Picture_6_1}
\end{figure}

\begin{figure}[ht]
\centering
\includegraphics[width=3.7326in,height=1.5in]{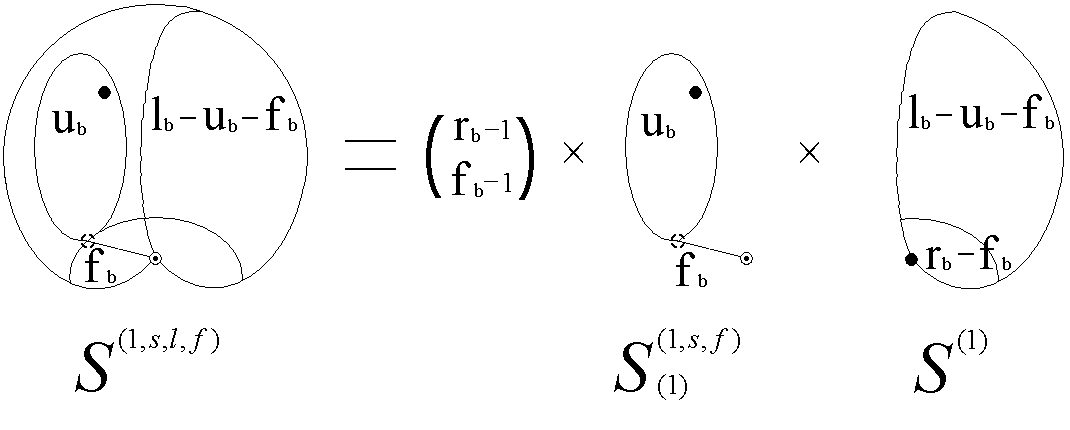}
\caption{Representation of  ${}^{(1)}\mathrm{Set}^{(1,\,s,\,l,\,f)}$}\label{Picture_7_1}
\end{figure}

\begin{figure}[ht]
\centering
\includegraphics[width=5.4625in,height=1.5in]{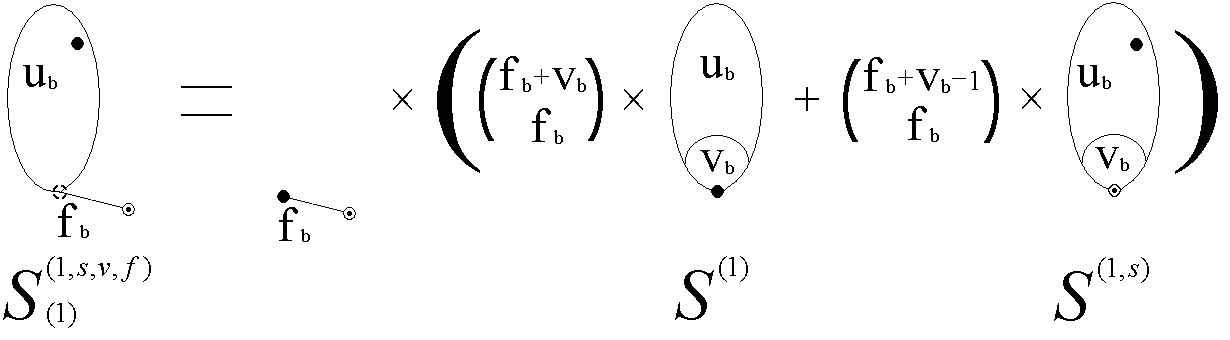}
\caption{Representation of $\mathrm{Set}^{(1,\,s,\,v,\,f)}_{(1)}$}\label{Picture_7_2}
\end{figure}

In a walk from $\mathrm{S}_{(\not =,\, \not c)}^{(s)}(u_g,u_b;v_g,v_b)$
 either there are gray edges with a gray root or there are not.
 If the walk has gray edges with a gray root, then the edge $(r,v)$ can be either blue-gray or
 pure gray.
If the edge $(r,v)$ is blue-gray, then it is just tree-like
 double walk with different roots and a blue-gray edge $ (r, v) $, i.e.
 she is from $\mathrm{S}_{(\not =,\, c)}^{(r)}(l_g,l_b;r_g,r_b)$.
If the edge $(r,v)$ is pure gray, then the blue root is in the lower graph, since
 a blue walk passes through the gray root  $r$. If there are no gray edges in the walk with a gray root, then there are no gray edges at all.
 \begin{multline}\label{rec_sn=nc}
 {}^{(1)}\mathrm{S}_{(\not =,\, \not c)}^{(s)}(l_g,l_b;r_g,r_b)=
{}^{(1)} \mathrm{S}_{(\not =,\, c)}^{(r)}(l_g,l_b;r_g,r_b) +
\\
 +{}^{(1)}\mathrm{S}_{(\not =,\, \not c)}^{(s,\, g,\,d)}(l_g,l_b;r_g,r_b)+
 {}^{(1)}\mathrm{S}_{(\not =,\, \not c)}^{(s,n)}(l_g,l_b;r_g,r_b)
\end{multline}
\begin{multline}\label{rec_sn=nc_1}
 {}^{(2)}\mathrm{S}_{(\not =,\, \not c)}^{(s)}(l_g,l_b;r_g,r_b)=
 {}^{(2)}\mathrm{S}_{(\not =,\, c)}^{(r)}(l_g,l_b;r_g,r_b) +
 \\
+{}^{(2)}\mathrm{S}_{(\not =,\, \not c)}^{(s,\, g,\,d)}(l_g,l_b;r_g,r_b)+
 {}^{(2)}\mathrm{S}_{(\not =,\, \not c)}^{(s,n)}(l_g,l_b;r_g,r_b)
\end{multline}
And we have the next lemma for $\mathrm{S}_{(\not =,\,\not c)}^{(s,\,g,\, d)}(l_g,l_b;r_g,r_b)$ (see also Fig. \ref{Picture_6_1}).
\begin{lemma}\label{l3.12} Let $l_g$, $l_b$, $r_g$, $r_b$ be natural numbers or zero such that $l_g \geq r_g > 0$ and $l_b \geq r_b > 0$. Then the following equalities are true.
 $$
 {}^{(1)}\mathrm{S}_{(\not =,\, \not c)}^{(s,\,g,\, d)}(l_g,l_b;r_g,r_b)\!=\!\sum_{u_g=0}^{l_g-r_g}\sum_
 {f_g=1}^{r_g}{}^{(1)}\mathrm{S}^{(s,\,g,\,d\,,l,\,f)}_{(\not =,\, \not c)}(l_g,l_b;r_g,r_b;u_g,f_g)
$$
$$
{}^{(1)}\mathrm{S}^{(s,\,g,\,d\,,l,\,f)}_{(\not =,\, \not c)}(l_g,l_b;r_g,r_b;u_g,f_g)\!=\!
\alpha_1^{-1}\cdot \binom{r_g-1}{f_g-1}\cdot
{}^{(1)}\mathrm{S}^{(1)}_{(1)}(f_g+u_g,f_g) \cdot {}^{(1)}\mathrm{S}_{(\not =,\,
\not c)}^{(s)}(l_g-u_g-f_g,l_b,r_g-f_g,r_b)
$$
$$
 {}^{(2)}\mathrm{S}_{(\not =,\, \not c)}^{(s,\,g,\, d)}(l_g,l_b;r_g,r_b)\!=\!\sum_{u_g=0}^{l_g-r_g}\sum_
 {f_g=1}^{r_g}{}^{(1)}\mathrm{S}^{(s,\,g,\,d\,,l,\,f)}_{(\not =,\, \not c)}(l_g,l_b;r_g,r_b;u_g,f_g)
$$
$$
{}^{(2)}\mathrm{S}^{(s,\,g,\,d\,,l,\,f)}_{(\not =,\, \not c)}(l_g,l_b;r_g,r_b;u_g,f_g)\!=\!
\alpha_2^{-1}\cdot\binom{r_g-1}{f_g-1}\cdot
{}^{(2)}\mathrm{S}^{(1)}_{(1)}(f_g+u_g,f_g) \cdot {}^{(1)}\mathrm{S}_{(\not =,\,
\not c)}^{(s)}(l_g-u_g-f_g,l_b,r_g-f_g,r_b)
$$
\end{lemma}
 Here we use lemma \ref{l2}.
$$
{}^{(1)}\mathrm{S}^{(1)}_{(1)}(f_g+u_g,f_g)=\sum_{v_g=0}^{u_g}
\binom{f_g+v_g-1}{f_g-1}\cdot \frac{\alpha_1\cdot V_{2f_g}}{p^{f_g-1}}
\cdot {}^{(2)}\mathrm{S}^{(1)}(u_g,v_g)
$$
$$
{}^{(2)}\mathrm{S}^{(1)}_{(1)}(f_g+u_g,f_g)=\sum_{v_g=0}^{u_g}
\binom{f_g+v_g-1}{f_g-1}\cdot \frac{\alpha_2\cdot V_{2f_g}}{p^{f_g-1}}
\cdot {}^{(1)}\mathrm{S}^{(1)}(u_g,v_g)
$$
Therefore, the following equalities hold.
$$
 {}^{(1)}\mathrm{S}_{(\not =,\, c)}^{(s,\,g,\, d)}(l_g,l_b;r_g,r_b)=\sum_{f_g=1}^{r_g} \binom{r_g-1}{f_g-1} \cdot
         \frac{ V_{2f_g}}{p^{f_g-1}}
        \cdot \sum_{u_g=0}^{l_g-r_g} {}^{(1)}\mathrm{S}_{(\not =,\, \not c)}^{(s)}(l_g-u_g-f_g,l_b;r_g-f_g,r_b) \cdot
$$
\begin{equation}\label{rec_sgdn=c}
\cdot \sum_{v_g=0}^{u_g} \binom{f_g+v_g-1}{f_g-1} \cdot
{}^{(2)}\mathrm{S}^{(1)}(u_g,v_g).
\end{equation}
$$
 {}^{(2)}\mathrm{S}_{(\not =,\, c)}^{(s,\,g,\, d)}(l_g,l_b;r_g,r_b)=\sum_{f_g=1}^{r_g} \binom{r_g-1}{f_g-1} \cdot
         \frac{ V_{2f_g}}{p^{f_g-1}}
        \cdot \sum_{u_g=0}^{l_g-r_g} {}^{(2)}\mathrm{S}_{(\not =,\, \not c)}^{(s)}(l_g-u_g-f_g,l_b;r_g-f_g,r_b) \cdot
$$
\begin{equation}\label{rec_sgdn=c_1}
\cdot \sum_{v_g=0}^{u_g} \binom{f_g+v_g-1}{f_g-1} \cdot
{}^{(1)}\mathrm{S}^{(1)}(u_g,v_g).
\end{equation}
If there are no gray edges, then $l_g\!=\!0$ and $r_g\!=\!0$.
So by definition of ${}^{(1)}\mathrm{S}_{(\not =,\, \not c)}^{(s,n)}(l_g,l_b;r_g,r_b)$ the following formulas are correct
 \begin{equation}\label{snn=nc}
{}^{(1)}\mathrm{S}_{(\not =,\, \not c)}^{(s,n)}(l_g,l_b;r_g,r_b)=\delta_{l_g}\cdot \delta_{r_g} \cdot  {}^{(1)}\mathrm{S}^{(1,s)}(l_b;r_b),
\end{equation}
 \begin{equation}\label{snn=nc_1}
{}^{(2)}\mathrm{S}_{(\not =,\, \not c)}^{(s,n)}(l_g,l_b;r_g,r_b)=\delta_{l_g}\cdot \delta_{r_g} \cdot  {}^{(2)}\mathrm{S}^{(1,s)}(l_b;r_b).
\end{equation}
Case $\mathrm{S}^{(1,\,s)}(l_b,r_b)$ also does not cause
problems (see fig. \ref{Picture_7_1} and \ref{Picture_7_2}). We denote by $b$ the vertex from which the blue walk first falls into the gray root  $r$. We cut along the vertices of the edge $(r,b)$. $u_b$ here means half of the length of  blue  walk along upper blue graph
 which was formed after removing the edge $ (r, b) $,
 and $ f_b $ means half of the length of the blue walk along the edge of $(r,b)$. By
$v_b$ we denote the number of steps leaving the vertex $b$ other than
$\overrightarrow {(b,r)}$. Obviously, the blue root lies in the upper graph. Again, two cases are possible: the blue root coincides with the vertex $b$ or it doesn't coincide with the vertex $b$. In different cases
different factors appear. And we have two more lemmas.
\begin{lemma}\label{l3.13} Let $l_b, r_b$ be natural numbers  such that  $l_b\geq r_b> 0$. Then the following equalities are true.
$$
 {}^{(1)}\mathrm{S}^{(1,\,s)}(l_b,r_b))\!=\!\sum_{u_b=0}^{l_b-r_b}\sum_
 {f_b=1}^{r_b}{}^{(1)}\mathrm{S}^{(1,\,s,\,l,\,f)}(l_b;r_b;u_b;f_b)
$$
$$
{}^{(1)}\mathrm{S}^{(1,\,s,\,l,\,f)}(l_b;r_b;u_b;f_b)\!=\!\binom{r_b-1}{f_b-1}\cdot \alpha_1^{-1}\cdot
{}^{(1)}\mathrm{S}^{(1,\,s,\,f)}_{(1)}(f_b+u_b,f_b) \cdot
{}^{(1)}\mathrm{S}^{(1)}(l_b-u_b-f_b,r_b-f_b)
$$
$$
 {}^{(2)}\mathrm{S}^{(1,\,s)}(l_b,r_b))\!=\!\sum_{u_b=0}^{l_b-r_b}\sum_
 {f_b=1}^{r_b}{}^{(1)}\mathrm{S}^{(1,\,s,\,l,\,f)}(l_b;r_b;u_b;f_b)
$$
$$
{}^{(2)}\mathrm{S}^{(1,\,s,\,l,\,f)}(l_b;r_b;u_b;f_b)\!=\!\binom{r_b-1}{f_b-1}\cdot \alpha_2^{-1}\cdot
{}^{(2)}\mathrm{S}^{(1,\,s,\,f)}_{(1)}(f_b+u_b,f_b) \cdot
{}^{(2)}\mathrm{S}^{(1)}(l_b-u_b-f_b,r_b-f_b)
$$
\end{lemma}
\begin{lemma}\label{l3.14} Let $f_b$ be a  natural number, $u_b$ be  a  natural number or zero. Then the following equalities are true.
$$
{}^{(1)}\mathrm{S}^{(1,\,s,\,f)}_{(1)}(f_b+u_b,f_b)=\sum_{v_b=0}^{u_b}
{}^{(1)}\mathrm{S}^{(1,\,s,\,v,\,f)}_{(1)}(f_b+u_b,f_b,v_b)
$$
$$
{}^{(1)}\mathrm{S}^{(1,\,s,\,v,\,f)}_{(1)}(f_b+u_b,f_b,v_b)\!=\!
\frac{\alpha_1\cdot V_{2f_b}}{p^{f_b-1}}\cdot\left(
\binom{f_b+v_b-1}{f_b}\cdot
{}^{(1)}\mathrm{S}^{(1,\,s)}(u_b,v_b)+\binom{f_b+v_b}{f_b}\cdot
{}^{(1)}\mathrm{S}^{(1)}(u_b,v_b)\right)
$$
$$
{}^{(2)}\mathrm{S}^{(1,\,s,\,f)}_{(1)}(f_b+u_b,f_b)=\sum_{v_b=0}^{u_b}
{}^{(2)}\mathrm{S}^{(1,\,s,\,v,\,f)}_{(1)}(f_b+u_b,f_b,v_b)
$$
$$
{}^{(2)}\mathrm{S}^{(1,\,s,\,v,\,f)}_{(1)}(f_b+u_b,f_b,v_b)\!=\!
\frac{\alpha_2\cdot V_{2f_b}}{p^{f_b-1}}\cdot\left(
\binom{f_b+v_b-1}{f_b}\cdot
{}^{(2)}\mathrm{S}^{(1,\,s)}(u_b,v_b)+\binom{f_b+v_b}{f_b}\cdot
{}^{(2)}\mathrm{S}^{(1)}(u_b,v_b)\right)
$$
\end{lemma}
These two lemmas imply the following equalities.
$$
 {}^{(1)}\mathrm{S}^{(1,\,s)}(l_b,r_b)=\sum_{f_b=1}^{r_b} \binom{r_b-1}{f_b-1}  \cdot \frac{ V_{2f_b}}{p^{f_b-1}}
        \cdot \sum_{u_b=0}^{l_b-r_b} {}^{(1)}\mathrm{S}^{(1)}(l_b-u_b-f_b,r_b-f_b) \,\cdot
$$
\begin{equation}\label{rec_1s}
\cdot \sum_{v_b=0}^{u_b} \left( \binom{f_b+v_b}{f_b} \cdot
{}^{(2)}\mathrm{S}^{(1)}(u_b,v_b)
       + \binom{f_b+v_b-1}{f_b} \cdot {}^{(2)}\mathrm{S}^{(1,\,s)}(u_b,v_b)
       \right).
\end{equation}
$$
 {}^{(2)}\mathrm{S}^{(1,\,s)}(l_b,r_b)=\sum_{f_b=1}^{r_b} \binom{r_b-1}{f_b-1}  \cdot \frac{ V_{2f_b}}{p^{f_b-1}}
        \cdot \sum_{u_b=0}^{l_b-r_b} {}^{(1)}\mathrm{S}^{(2)}(l_b-u_b-f_b,r_b-f_b) \,\cdot
$$
\begin{equation}\label{rec_1s1}
\cdot \sum_{v_b=0}^{u_b} \left( \binom{f_b+v_b}{f_b} \cdot
{}^{(1)}\mathrm{S}^{(1)}(u_b,v_b)
       + \binom{f_b+v_b-1}{f_b} \cdot {}^{(1)}\mathrm{S}^{(1,\,s)}(u_b,v_b)
       \right).
\end{equation}
 The lemmas for ${}^{(1)}\mathrm{S}_{(\not =,\, c)}^{(r,\, d)}(l_b,r_b)$ are formulated as follows.
 \begin{lemma}\label{l3.15} Let $l_g$, $l_b$, $r_g$, $r_b$ be natural numbers or zero such that $l_g \geq r_g > 0$ and $l_b \geq r_b > 0$. Then the following equalities are true.
 $$
 {}^{(1)}\mathrm{S}_{(\not =,\, c)}^{(r,\, d)}(l_g,l_b,r_g,r_b)\!=\!\sum_{u_g=0}^{l_g-r_g}\sum_
 {f_g=1}^{r_g}\sum_{u_b=0}^{l_b-r_b}\sum_
 {f_b=1}^{r_b}{}^{(1)}\mathrm{S}_{(\not =,\, c)}^{(r,\,d,\,l,\,f)}(l_g,l_b;r_g,r_b;u_g,u_b;f_g,f_b)
$$
$$
 {}^{(2)}\mathrm{S}_{(\not =,\, c)}^{(r,\, d)}(l_g,l_b,r_g,r_b)\!=\!\sum_{u_g=0}^{l_g-r_g}\sum_
 {f_g=1}^{r_g}\sum_{u_b=0}^{l_b-r_b}\sum_
 {f_b=1}^{r_b}{}^{(2)}\mathrm{S}_{(\not =,\, c)}^{(r,\,d,\,l,\,f)}(l_g,l_b;r_g,r_b;u_g,u_b;f_g,f_b)
$$
 $$
 {}^{(2)}\mathrm{S}_{(\not =,\,c)}^{(r,\,d,\,l,\,f)}(l_g,l_b;r_g,r_b;u_g,u_b;f_g,f_b)\!=\!
\binom{r_g-1}{f_g-1} \cdot\binom{r_b-1}{f_b}\cdot
 {}^{(2)}\mathrm{S}_{(1,\, =,\, c)}(f_g+u_g,f_b+u_b;f_g,f_b) \cdot
$$
$$
\alpha_2^{-1}\cdot {}^{(2)}\mathrm{S}^{(s)}_{(\not =,\, \not c)}
(l_g-u_g-f_g,l_b-u_b-f_b;r_g-f_g,r_b-f_b)
$$
 \end{lemma}
\begin{lemma}\label{l3.16}
Let $l_g$, $l_b$, $r_g$, $r_b$, $f_g$, $f_b$ be natural numbers and $u_g$, $v_g$ be natural numbers or zero such that $l_g \geq r_g \geq f_g > 0$ and $l_b \geq r_b \geq f_b > 0$. Then the following equalities are true.
$$
 {}^{(1)}\mathrm{S}_{(\not =,\,
c)}^{(r,\,d,\,l,\,f)}(l_g,l_b;r_g,r_b;u_g,u_b;f_g,f_b)\!=\!
\binom{r_g-1}{f_g-1} \cdot\binom{r_b-1}{f_b}\cdot
 {}^{(1)}\mathrm{S}_{(1,\, =,\, c)}(f_g+u_g,f_b+u_b;f_g,f_b) \cdot
$$
$$
\alpha_1^{-1}\cdot {}^{(1)}\mathrm{S}^{(s)}_{(\not =,\, \not c)}
(l_g-u_g-f_g,l_b-u_b-f_b;r_g-f_g,r_b-f_b)
$$
 $$
   {}^{(1)}\mathrm{S}_{(1,\,=,\,c)}(f_g+u_g,f_b+u_b;f_g,f_b)=\sum_{v_g=0}^{u_g}\sum_{v_b=0}^{u_b}
 {}^{(1)}\mathrm{S}^{(v)}_{(1,\, =,\, c)}(f_g+u_g,f_b+u_b;f_g,f_b;v_g,v_b)
$$
$$
  {}^{(1)}\mathrm{S}^{(v)}_{(1,\, =,\,c)}(f_g+u_g,f_b+u_b;f_g,f_b;v_g,v_b)\!=\!
\binom{f_g+v_g-1}{f_g-1}\cdot \binom{f_b+v_b-1}{f_b-1} \, \cdot
$$
$$
 \alpha_1\cdot  \frac{V_{2\cdot(f_g+f_b)}}{p^{f_g+f_b-1}} \cdot{}^{(2)}\mathrm{S}_{( =,\,
\not c)}(u_g,u_b;v_g,v_b).
$$
 $$
   {}^{(2)}\mathrm{S}_{(1,\,=,\,c)}(f_g+u_g,f_b+u_b;f_g,f_b)=\sum_{v_g=0}^{u_g}\sum_{v_b=0}^{u_b}
 {}^{(2)}\mathrm{S}^{(v)}_{(1,\, =,\, c)}(f_g+u_g,f_b+u_b;f_g,f_b;v_g,v_b)
$$
$$
  {}^{(2)}\mathrm{S}^{(v)}_{(1,\, =,\,c)}(f_g+u_g,f_b+u_b;f_g,f_b;v_g,v_b)\!=\!
\binom{f_g+v_g-1}{f_g-1}\cdot \binom{f_b+v_b-1}{f_b-1} \, \cdot
$$
$$
 \alpha_2 \cdot   \frac{V_{2\cdot(f_g+f_b)}}{p^{f_g+f_b-1}} \cdot {}^{(1)}\mathrm{S}_{( =,\,
\not c)}(u_g,u_b;v_g,v_b).
$$
\end{lemma}
These two lemmas \ref{l3.15} and \ref{l3.16} imply the following equalities.
$$
  {}^{(1)}\mathrm{S}_{(\not =,\, c)}^{(r,\, d)}(l_g,l_b;r_g,r_b)=\sum_{f_g=1}^{r_g} \binom{r_g-1}{f_g-1}
  \cdot \sum_{f_b=1}^{r_b}
       \binom{r_b-1}{f_b}
       \frac{ V_{2(f_g+f_b)}}{p^{f_g+f_b-1}} \cdot
$$
$$
 \cdot      \sum_{u_g=0}^{l_g-r_g}
       \sum_{u_b=0}^{l_b-r_b}  {}^{(1)}\mathrm{S}_{(\not =,\, \not c)}^{(s)}
       (l_g-u_g-f_g,l_b-u_b-f_b;r_g-f_g,r_b-f_b) \cdot
$$
\begin{equation}\label{rec_rdn=c}
\cdot  \sum_{v_g=0}^{u_g} \binom{f_g+v_g-1}{f_g-1} \cdot
      \sum_{v_b=0}^{u_b} \binom{f_b+v_b-1}{f_b-1} \cdot  {}^{(2)}\mathrm{S}_{( =,\, \not
      c)}(u_g,u_b;v_g,v_b).
\end{equation}
$$
  {}^{(2)}\mathrm{S}_{(\not =,\, c)}^{(r,\, d)}(l_g,l_b;r_g,r_b)=\sum_{f_g=1}^{r_g} \binom{r_g-1}{f_g-1}
  \cdot \sum_{f_b=1}^{r_b}
       \binom{r_b-1}{f_b}
       \frac{ V_{2(f_g+f_b)}}{p^{f_g+f_b-1}} \cdot
$$
$$
 \cdot      \sum_{u_g=0}^{l_g-r_g}
       \sum_{u_b=0}^{l_b-r_b}  {}^{(2)}\mathrm{S}_{(\not =,\, \not c)}^{(s)}
       (l_g-u_g-f_g,l_b-u_b-f_b;r_g-f_g,r_b-f_b) \cdot
$$
\begin{equation}\label{rec_rdn=c_1}
\cdot  \sum_{v_g=0}^{u_g} \binom{f_g+v_g-1}{f_g-1} \cdot
      \sum_{v_b=0}^{u_b} \binom{f_b+v_b-1}{f_b-1} \cdot  {}^{(1)}\mathrm{S}_{( =,\, \not
      c)}(u_g,u_b;v_g,v_b).
\end{equation}

The system $\ref{rec}-\ref{rec_=c}$, $\ref{rec_g=c}-\ref{rec_rdn=c}$ is recursive, since by
each essential step we decrease the total length of the double
(single) walk. For unambiguous solvability, it is necessary to impose on the system
the following initial conditions.
\begin{equation}\label{ini_1}
{}^{(1)}\mathrm{S}^{(1)}(0,x)\!=\!\delta_x\cdot \alpha_1, {}^{(2)}\mathrm{S}^{(1)}(0,x)\!=\!\delta_x\cdot \alpha_2
\end{equation}
\begin{equation}\label{ini_2}
   {}^{(1)}\mathrm{S}^{(1,\,s)}(0,x)\!=\!0, {}^{(2)}\mathrm{S}^{(1,\,s)}(0,x)\!=\!0
\end{equation}
 Let $x, y, z, w$ be natural numbers or zero such that at least one of the next inequalities $z\geq x>0$ $w\geq y>0$ is violated. Then the following equalities hold.
\begin{equation}\label{ini_3}
{}^{(1)}\mathrm{S}_{(=,\,  c)}^{(g)}(z,w;x,y)\!=\!0, {}^{(2)}\mathrm{S}_{(=,\,  c)}^{(g)}(z,w;x,y)\!=\!0
\end{equation}
\begin{equation}\label{ini_4}
 {}^{(1)}\mathrm{S}_{(=,\,  c)}^{(r)}(z,w;x,y)\!=\!0,  {}^{(2)}\mathrm{S}_{(=,\,  c)}^{(r)}(z,w;x,y)\!=\!0
\end{equation}
\begin{equation}\label{ini_5}
  {}^{(1)}\mathrm{S}_{(\not =,\, c)}^{(r,\, u)}(z,w;x,y)\!=\!0,  {}^{(2)}\mathrm{S}_{(\not =,\, c)}^{(r,\, u)}(z,w;x,y)\!=\!0
\end{equation}
\begin{equation}\label{ini_6}
  {}^{(1)}\mathrm{S}_{(\not =,\, c)}^{(r,d)}(z,w;x,y)\!=\!0, {}^{(2)}\mathrm{S}_{(\not =,\, c)}^{(r,d)}(z,w;x,y)\!=\!0
\end{equation}
\begin{equation}\label{ini_7}
   {}^{(1)}\mathrm{S}_{(\not =,\, \not c)}^{(s,\,g,\, d)}(z,w;x,y)\!=\!0, {}^{(2)}\mathrm{S}_{(\not =,\, \not c)}^{(s,\,g,\, d)}(z,w;x,y)\!=\!0
\end{equation}
 Let $x, y, z, w$ be natural numbers or zero such that at least one of the next inequalities $z\geq x>0$ $w\geq y\geq 0$ is violated. Then the following equalities hold.
\begin{equation}\label{ini_8}
  {}^{(1)}\mathrm{S}_{(\not =,\, c)}^{(g,\, u)}(z,w;x,y)\!=\!0, {}^{(2)}\mathrm{S}_{(\not =,\, c)}^{(g,\, u)}(z,w;x,y)\!=\!0
\end{equation}
\begin{equation}\label{ini_9}
{}^{(1)}\mathrm{S}_{(\not =,\, c)}^{(g,d)}(z,w;x,y)\!=\!0, {}^{(2)}\mathrm{S}_{(\not =,\, c)}^{(g,d)}(z,w;x,y)\!=\!0
\end{equation}





\begin{thebibliography}{99}

\bibitem{W} E.P.Wigner. On the distribution of the roots of
certain symmetric matrices, Ann.Math. {\bf 67}: (1958), 325-327.




\bibitem{KKPS}  A. Khorunzhy, B. Khoruzhenko, L. Pastur and
M. Shcherbina. The Large-n Limit in Statistical Mechanics and
Spectral Theory
     of Disordered Systems.
     Phase transition and critical phenomena.v.15, p.73, Academic
     Press, 1992

\bibitem{Pa:00} Pastur, L.: Random matrices as paradigm. In: Fokas,
A.,
Grigoryan, A., Kibble, T., Zegarlinski B. (eds.)
\emph{Mathematical Physics 2000}. London: Imperial College Press,
(2000), pp. 216-266.

\bibitem{RB:88} G.J. Rodgers  and A.J. Bray. Density of states of a
sparse random matrix, Phys.Rev.B {\bf 37}, (1988), 3557-3562.

\bibitem{RD:90}  G.J. Rodgers and C. De Dominicis. Density of states of
sparse random matrices,
J.Phys.A:Math.Jen.{\bf 23}, (1990), 1567-1566.

\bibitem{MF:91} A.D.Mirlin, Y.V.Fyodorov. Universality of the
level correlation function of sparce random matrices,
  J.Phys.A:Math.Jen.{\bf 24}, (1991), 2273-2286.

\bibitem{FM:96} Y.V.Fyodorov, A.D.Mirlin. Strong eigenfunction
correlations near the Anderson localization transition.
arXiv:cond-mat/9612218 v1


\bibitem{BG1} M.Bauer and O.Golinelli. Random incidence matrices:
spectral density at zero energy, Saclay preprint T00/087;
cond-mat/0006472

\bibitem{BG2} M.Bauer and O.Golinelli. Random incidedence matrices:
moments and
spectral density, J.Stat. Phys. {\bf 103}, 301-336, 2001


\bibitem{B} B. Bollobas {\it Random Graphs }   Acad. Press (1985)




























\bibitem{KV}  A. Khorunzhy, V. Vengerovsky. On asymptotic solvability
of random graph's laplacians. Preprint \textit{lanl.arXiv.org}
math-ph/0009028

\bibitem{KSV} Khorunzhy O., Shcherbina M., and Vengerovsky V. Eigenvalue distribution of large weighted random graphs, J. Math. Phys. {\bf 45}  N.4: (2004), 1648-1672.



\bibitem{Me:91} M.L.Mehta: \emph{Random Matrices}. New York: Academic
Press, 1991







\bibitem{V0} V. Vengerovsky. Asymptotics of the correlator of some ensemble of sparse random matrices,
JMPAG {\bf 04-2}, 2016, 135-160.


\bibitem{V} V. Vengerovsky. Eigenvalue Distribution of a Large Weighted Bipartite Random Graph,
JMPAG {\bf 10-2}, 2014, 240-255.

\bibitem{V1} V. Vengerovsky. Eigenvalue Distribution of a Large Weighted Bipartite Random Graph. Resolvent Approach,
JMPAG {\bf 12-1}, 2016, 78-93.



\bibitem{KKP} {\it A. Khorunzhy, B. Khoruzhenko, and L. Pastur},
{\textnormal Asymptotic properties of large random matrices with independent
entries}.--- {\it J. Math. Phys.}{\textnormal(1996), v. 37, p.
5033-5059.}

\bibitem{APS2} {\it S. Albeverio, L. Pastur, M. Shcherbina},{\textnormal On the $1/n$
expansion for some unitary invariant ensembles of random
matrices}.---
{\it Commun. Math. Phys.}{\textnormal (2001), v. 224,  p. 271-305.}

\bibitem{PJ} {\it Paul Jung, Jaehun Lee},
{\textnormal Delocalization and limiting spectral distribution of
Erdos-Renyi graphs with constant expected degree}.--- {\it Electron. Commun. Probab. }  {\textnormal (2018), v. 23,  p. 1-13.}

\bibitem{CS} {\it Simon Coste, Justin Salez},
{\textnormal Emergence of extended states at zero in the spectrum of sparse random graphs}.--- {\it The Annals of Probability }  {\textnormal (2021), v.  49 (4),  p. 2012 - 2030.}
\end{thebibliography}
\end{document}